\documentclass{emulateapj}
\usepackage{verbatim,graphicx,natbib}
\shortauthors{Cowan et al.}
\shorttitle{Rotational Variability of Earth's Polar Regions}

\begin{document}
\title{Rotational Variability of Earth's Polar Regions: Implications for Detecting Snowball Planets}
\author{Nicolas B. Cowan\altaffilmark{1,2}, Tyler Robinson\altaffilmark{3,4}, Timothy A. Livengood\altaffilmark{5}, Drake Deming\altaffilmark{5,4}, Eric Agol\altaffilmark{3}, Michael F. A'Hearn\altaffilmark{6}, David Charbonneau\altaffilmark{7}, Carey M. Lisse\altaffilmark{8}, Victoria S. Meadows\altaffilmark{3,4}, Sara Seager\altaffilmark{9,4}, Aomawa L.\ Shields\altaffilmark{3,4}, Dennis D. Wellnitz\altaffilmark{6}}
\altaffiltext{1}{Northwestern University, 2131 Tech Drive, Evanston, IL 60208\\ Email: n-cowan@northwestern.edu}
\altaffiltext{2}{CIERA Postdoctoral Fellow}
\altaffiltext{3}{Astronomy Department \& Astrobiology Program, University of Washington, Box 351580, Seattle, WA  98195}
\altaffiltext{4}{NASA Astrobiology Institute Member}
\altaffiltext{5}{NASA Goddard Space Flight Center, Greenbelt, MD 20771}
\altaffiltext{6}{Department of Astronomy, University of Maryland, College Park MD 20742}
\altaffiltext{7}{Harvard-Smithsonian Center for Astrophysics, 60 Garden Street, Cambridge, MA 02138}
\altaffiltext{8}{Johns Hopkins University Applied Physics Laboratory, SD/SRE, MP3-E167, 11100 Johns Hopkins Road, Laurel, MD 20723}
\altaffiltext{9}{Department of Earth, Atmospheric, and Planetary Sciences, Dept of Physics, Massachusetts Institute of Technology, 77 Massachusetts Ave. 54-1626, MA 02139}

\begin{abstract}
We have obtained the first time-resolved, disc-integrated observations of Earth's poles with the Deep Impact spacecraft as part of the EPOXI Mission of Opportunity. These data mimic what we will see when we point next-generation space telescopes at nearby exoplanets. We use principal component analysis (PCA) and rotational lightcurve inversion to characterize color inhomogeneities and map their spatial distribution from these unusual vantage points, as a complement to the equatorial views presented in \cite{Cowan_2009}. We also perform the same PCA on a suite of simulated rotational multi-band lightcurves from NASA's Virtual Planetary Laboratory 3D spectral Earth model. This numerical experiment allows us to understand what sorts of surface features PCA can robustly identify. We find that the EPOXI polar observations have similar broadband colors as the equatorial Earth, but with 20--30\% greater apparent albedo. This is because the polar observations are most sensitive to mid-latitudes, which tend to be more cloudy than the equatorial latitudes emphasized by the original EPOXI Earth observations.  The cloudiness of the mid-latitudes also manifests itself in the form of increased variability at short wavelengths in the polar observations, and as a dominant gray eigencolor in the south polar observation. We construct a simple reflectance model for a snowball Earth. By construction, our model has a higher Bond albedo than the modern Earth; its surface albedo is so high that Rayleigh scattering does not noticeably affect its spectrum. The rotational color variations occur at short wavelengths due to the large contrast between glacier ice and bare land in those wavebands. Thus we find that both the broadband colors and diurnal color variations of such a planet would be easily distinguishable from the modern-day Earth, regardless of viewing angle.
 \end{abstract}

\section{Introduction}
Observations of exoplanets will be limited to disc-integrated measurements for the foreseeable future. This is true whether a planet can be spatially resolved from its host star (direct imaging) or not (as in current studies of short-period transiting planets). Spectra with long integration times yield invaluable information about the spatially and temporally averaged composition and temperature-pressure profile of the atmosphere. 

Time-resolved photometry, on the other hand, tells us about the weather, climate, and spatial inhomogeneities of the planet. The time-variability of a planet occurs on two timescales, rotational and orbital\footnote{These are one and the same for synchronously rotating planets.}, and yields different information depending on whether it is observed in reflected or thermal light.  

\emph{Thermal phases} inform us about the diurnal heating patterns of the planet: the day-side temperature, the night-side temperature, and the hottest local time on the planet \citep{Cowan_2008}. Depending on whether it has an atmosphere, such observations can constrain a body's rotation rate, as well as its average Bond albedo, thermal inertia, emissivity, surface roughness and wind velocities \citep{Spencer_1990, Cowan_2011}.

\emph{Rotational variations in thermal emission} are caused by inhomogeneities in the planet's albedo and thermal inertia. This has been studied for minor solar system bodies, where it can be used to break the degeneracy between albedo markings and shape \citep[eg,][]{Lellouch_2000}.

\emph{Reflected phases} are a measure of the disk-integrated scattering phase function, telling us ---for example--- about clouds and oceans on the planet, especially when combined with polarimetry \citep{Williams_2008, Mallama_2009, Robinson_2010, Zugger_2010}.

\emph{Rotational variations at reflected wavelengths} can identify the rotation rate of an unresolved planet \citep{Palle_2008}. Once the rotation rate has been determined, one can constrain the albedo markings on a world \citep{Russell_1906}, indicating surface features like continents and oceans \citep{Ford_2001, Cowan_2009, Oakley_2009, Fujii_2010}. Finally, the spatial distribution of landmasses can be inferred, and the planet's obliquity can be estimated if diurnal variations are monitored at a variety of phases \citep{Cowan_2009, Oakley_2009, Kawahara_2010}. 

In this paper, we study rotational (a.k.a.\ diurnal) variability of Earth's poles at visible wavelengths. At these wavelengths, the observed flux consists entirely of reflected sunlight. Earthshine ---the faint illumination of the dark side of the Moon due to reflected light from Earth--- was first explaied in the early 16$^{\rm th}$ century by Leonardo Da Vinci, and has been used more recently to study the reflectance spectrum and cloud cover variability \citep{Goode_2001, Qiu_2003, Palle_2003, Palle_2004, Montanes_2007}, vegetation ``red edge'' signature \citep{Woolf_2002, Montanes_2005, Seager_2005, Hamdani_2006, Montanes_2006} and the effects of specular reflection \citep{Woolf_2002, Langford_2009} for limited regions of our planet. More recently, \cite{Palle_2009} measured the disc-integrated \emph{transmission} spectrum of Earth by observing the Moon during a lunar eclipse. Brief snapshots of Earth obtained with the Galileo spacecraft have been used to study our planet from afar \citep{Sagan_1993, Geissler_1995} and numerical models have been developed to anticipate how diurnal variations in disc-integrated light could be used to characterize Earth \citep[][Robinson et al.\ 2011]{Ford_2001, Tinetti_2006, Tinetti_2006b, Palle_2008, Williams_2008, Oakley_2009, Fujii_2010, Kawahara_2010, Zugger_2010, Robinson_2010}.

In \cite{Cowan_2009} we presented two epochs of rotational variability for disk-integrated Earth seen equator-on. We performed a principal component analysis (PCA) of the 7 optical wavebands to identify the eigencolors of the variability. 
There were two components that accounted for more than 98\% of the color variations seen.  The two-dimensionality of the PCA indicated that three major surface types were necessary to explain the observed variability.  The dominant eigencolor was red, which we identified as being primarily sensitive to cloud-free land.  A rotational inversion of the red eigenprojection yielded a rough map of the major landforms of Earth: the Americas, Africa-Eurasia and Oceania, separated by the major oceans: the Atlantic and Pacific.

One concern with the diurnal light curve inversion of \cite{Cowan_2009} and \cite{Oakley_2009} is the unknown obliquity of the planet: there is no good Bayesian prior for the obliquities and rotation rates, except that they will be slightly biased towards prograde rotation \citep{Schlichting_2007}. Numerical experiments have shown that a pole-on viewing geometry might complicate retrieval of a planet's rotational period \citep{Palle_2008}, but once that periodicity is identified, it is possible to create albedo maps of the planet, although without knowledge of the planet's obliquity one will not know what latitudes those maps correspond to \citep{Cowan_2009}. Idealized numerical experiments show that ---in principle--- the obliquity can be extracted if one observes diurnal variability at a variety of phases \citep{Kawahara_2010}, but it is not yet clear how well such a technique would work for a cloudy planet with unknown surface types. 

Finaly, Earth's climate is sensitive to the latitudinal configuration of continents and few percent changes in insolation (neither of which will be well constrained for exoplanets) leading to bifurcations between temperate and snowball climates \citep{Voigt_2010}.  Given an extrasolar Earth analog, how can we use optical photometry to distinguish between the two branches of this positive feedback loop?     

In this paper we report and analyze disk-integrated observations of Earth's polar regions obtained from the Deep Impact spacecraft as part of the EPOXI mission of opportunity. In \S~2 we present the observations; in \S~3 we discuss the color variability; in \S~4 we make longitudinal profiles of these colors; in \S~5 we introduce a simple model for snowball Earth and compare its diurnal color variations to those of Earth's polar regions; we summarize our results in \S~6 and state the implications of this study for mission planning in \S~7. 

\section{Observations}
The EPOXI\footnote{The University of Maryland leads the overall EPOXI mission, including the flyby of comet Hartley 2. NASA Goddard leads the exoplanet and Earth observations.} mission reuses the still-functioning Deep Impact spacecraft
that successfully observed comet 9P/Tempel~1. EPOXI science targets
include several transiting exoplanets and Earth \emph{en route} to a
flyby of comet 103P/Hartley~2. 

The first round of EPOXI Earth observations were taken from a vantage point very near the Earth's equatorial plane \citep[][Robinson et al. 2010a; Livengood et al. 2011]{Cowan_2009}. In the current paper, we focus on the later polar observations, summarized in Table~\ref{observations}. As with the equatorial observations, they were obtained with Earth near quadrature (phase angle $\alpha=\pi/2$), a favorable phase for directly imaging exoplanets, since the angular distance between the planet and its host star is maximized\footnote{Strictly speaking, this is only true for a planet on a circular orbit.}.

\begin{table*}[htb]
\begin{minipage}{126mm}
\caption{\bf EPOXI Earth Observations}
\label{observations}
\begin{tabular}{@{}llcccccc}
\hline
Name &Start of Observations& Starting & Sub-Observer& Sub-Solar& Dominant&Phase$^{3}$ & Illuminated Fraction\\
& & CML$^{1}$ & Latitude & Latitude&Latitude$^{2}$ && of Earth Disc$^{3}$\\ 
\hline
{\bf Earth1: Equinox} &2008 Mar 18, UTC 18:18:37 & $150^{\circ}$ W & $1.7^{\circ}$~N&$0.6^{\circ}$~S & 0.5$^{\circ}$~N&57$^{\circ}$ & 77\%\\
{\bf Earth5: Solstice} &2008 Jun 4, UTC 16:59:08 & $150^{\circ}$ W & $0.3^{\circ}$~N&$22.7^{\circ}$~N & 13$^{\circ}$~N&77$^{\circ}$ & 62\%\\
{\bf Polar1: North} &2009 Mar 27, UTC 16:19:42& $152^{\circ}$ W & $61.7^{\circ}$~N&$2.6^{\circ}$~N & $34^{\circ}$~N&87$^{\circ}$ & 53\%\\
{\bf Polar2: South} &2009 Oct 4, UTC 09:37:11 & $59^{\circ}$ W & $73.8^{\circ}$~S& $4.3^{\circ}$~S&$39^{\circ}$~S & 86$^{\circ}$ & 53\%\\
 \hline
\end{tabular}

\medskip
$^{1}$The CML is the Central Meridian Longitude, the longitude of the sub-observer point.\\ 
$^{2}$The dominant latitude is that expected to contribute the most photons, assuming a uniform Lambert sphere.\\ 
$^{3}$The planetary phase, $\alpha$, is the star--planet--observer angle and is related to the illuminated fraction by $f = \frac{1}{2}(1+\cos\alpha)$.
\end{minipage}
\end{table*}

Deep Impact's 30~cm diameter telescope coupled with the High Resolution Imager \citep[HRI,][]{Hampton_2005} recorded images of Earth in seven 100~nm wide optical
wavebands spanning 300--1000~nm, summarized in Table~\ref{bandpasses}.

\begin{table}[htb]
\caption{\bf EPOXI Photometric Bandpasses}
\label{bandpasses}
\begin{center}
\begin{tabular}{ccc}
\hline
Waveband & Cadence & Exp. Time\\ 
\hline
350~nm & 1 hr & 73.4~msec\\
450~nm & 15 min & 13.3~msec\\
550~nm & 15 min & 8.5~msec\\
650~nm & 15 min & 9.5~msec\\
750~nm & 1 hr & 13.5~msec\\
850~nm & 15 min & 26.5~msec\\
950~nm & 1 hr & 61.5~msec\\
 \hline
\end{tabular}
\end{center}
\end{table}

\begin{figure}[htb]
\includegraphics[width=84mm]{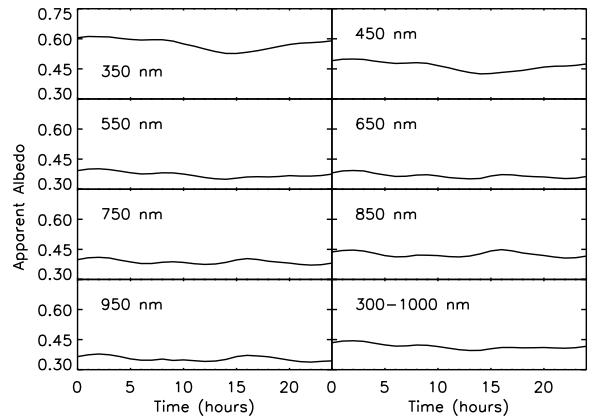}
\caption{{\bf North:} Light curves obtained in the seven HRI-VIS bandpasses by the EPOXI spacecraft when it passed above Earth's equatorial plane on March 27, 2009. The bottom-right panel shows changes in the bolometric albedo of Earth. The sub-observer longitude at the start of the observations is $152^{\circ}$W, the sub-observer latitude is $61.7^{\circ}$N throughout. The relative peak-to-trough variability ranges from 16\% (450~nm) to 10\% (750~nm).}
\label{polar1_lightcurves}
\end{figure}

\begin{figure}[htb]
\includegraphics[width=84mm]{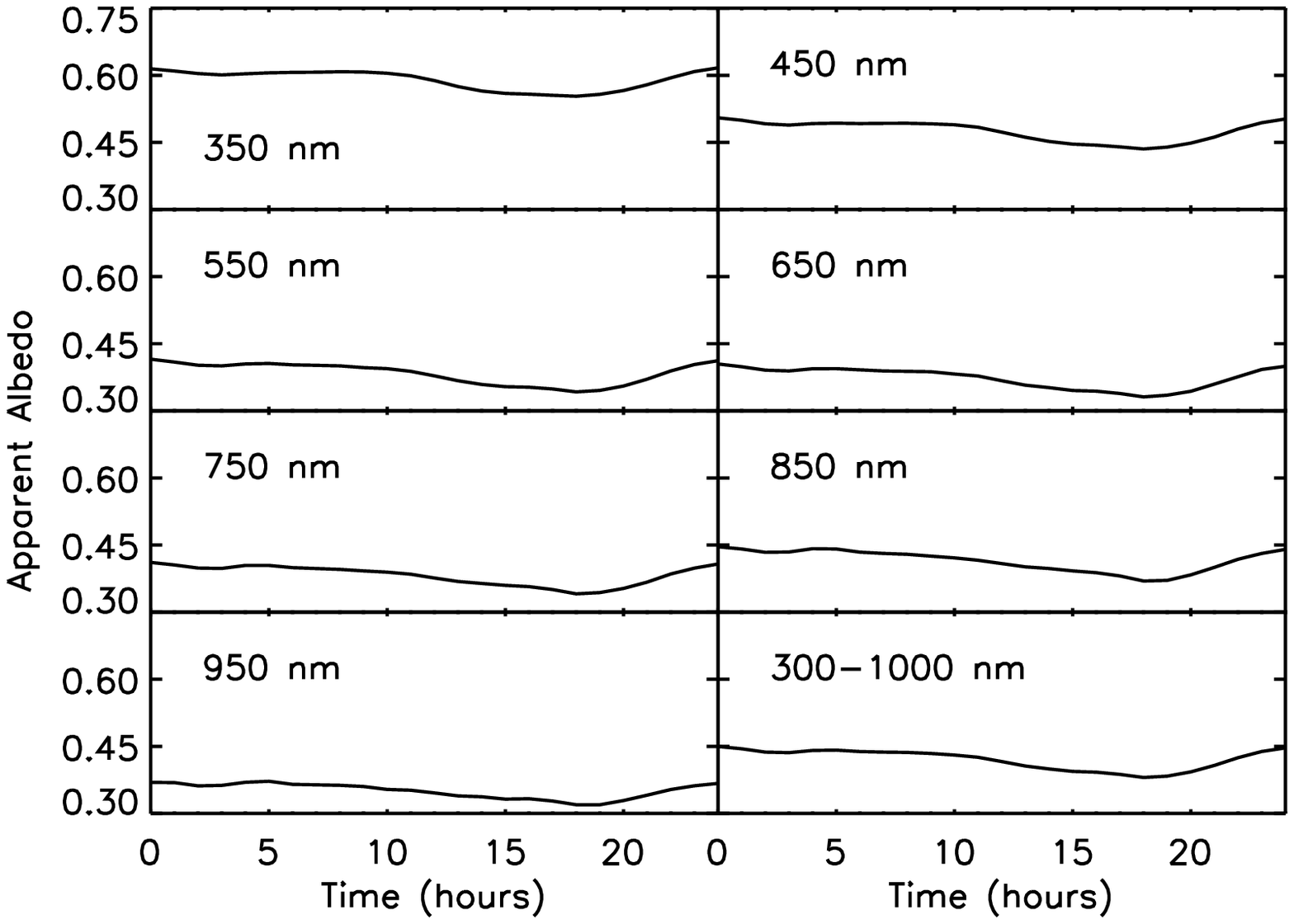}
\caption{{\bf South:} Light curves obtained in the seven HRI-VIS bandpasses by the EPOXI spacecraft when it passed below Earth's equatorial plane on October 4, 2009. The bottom-right panel shows changes in the bolometric albedo of Earth. The sub-observer longitude at the start of the observations is $59^{\circ}$W, the sub-observer latitude is $73.8^{\circ}$S  throughout. The relative peak-to-trough variability ranges from 11\% (350~nm) to 20\% (650~nm).}
\label{polar2_lightcurves}
\end{figure}

Although the EPOXI images of Earth offer spatial resolution of better than 100~km, we mimic the data that will eventually be
available for exoplanets by integrating the flux over the entire disc
of Earth and using only the hourly EPOXI observations from each of the wavebands, producing seven light curves for each of the two observing
campaigns, shown in
Figures~\ref{polar1_lightcurves} and \ref{polar2_lightcurves}. The photometric uncertainty in these data is exceedingly small: on the order of 0.1\% relative errors. 

Details of the observations and reduction will be presented in Livengood et al. (2011). After performing aperture photometry, the measured disk-integrated flux of Earth as seen from the Deep Impact spacecraft has units of specific flux [W m$^{-2}$ $\mu$m$^{-1}$]. We then apply the following steps: 1) We multiply by the HRI filter bandwidth of 0.1~$\mu$m to convert from specific flux to flux [W m$^{-2}$].  The detailed bandpass shapes are not important, provided that we use the same 0.1~$\mu$m top-hat bandpasses in computing the solar flux in step 3). 2) We divide the flux observed from the spacecraft by $(R_{\oplus}/r)^{2}$ to obtain the disc-averaged flux from Earth at the top of the atmosphere, where $R_{\oplus}$ is Earth's radius and $r$ is the spacecraft range. 3) We use a Kurucz model\footnote{Ideally, one would obtain a spectrum of the planet's host star with the same instrument used for imaging the planet, but this was not possible here for technical reasons.} for the solar specific flux at 1~AU, and convert it to flux in each of the HRI wavebands using the 0.1~$\mu$m top-hat bandpasses. 4) Dividing the result of steps 2) and 3) by each other, we obtain the top-of-the-atmosphere reflectance of the planet at the observed phase. 5) We further divide the reflectance by the scaled Lambert phase function \citep[$f(\alpha)=\frac{2}{3\pi}\left(\sin\alpha + (\pi-\alpha)\cos\alpha\right)$;][]{Russell_1916}, thus obtaining the planet's apparent albedo, $A^{*}$.  The precise definition of $A^{*}$ is given in Equation~5; for now it is sufficient to think of it as the average albedo of the planet, weighted by the illumination and visibility of various regions at that moment in time.

Unlike the observations presented in \cite{Cowan_2009}, the viewing geometry is sufficiently different for the two polar observations that we treat them separately. In particular, a single exoplanet could be observed by the same stationary observer with the two viewing geometries of \cite{Cowan_2009} since the sub-observer latitude was essentially equatorial at both epochs (see Table~1). Although the sub-stellar latitude varies with phase for non-zero obliquities, the sub-observer latitude is constant, provided that one can neglect precession (for more details on viewing geometry, see \S~\ref{viewing_geometry}).  By contrast,  the two time series presented in the current paper have sub-observer latitudes of   62$^{\circ}$~N and 74$^{\circ}$~S, respectively.

\begin{figure}[htb]
\includegraphics[width=84mm]{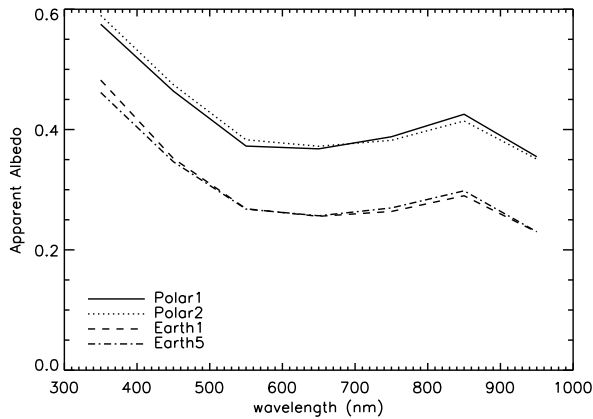}
\caption{24-hour average broadband spectra of Earth as seen from the Deep Impact spacecraft as part of the EPOXI mission.}
\label{average_albedo_spectrum}
\end{figure}

\begin{table*}[htb]
\begin{minipage}{126mm}
\caption{\bf EPOXI Earth Observations: Apparent Albedos}
\label{apparent_albedos}
\begin{tabular}{@{}lccccccc}
\hline
Name & 350~nm & 450~nm & 550~nm & 650~nm & 750~nm & 850~nm & 950~nm\\
\hline
{\bf Earth1: Equinox} & 0.482 (0.016)   &  0.352 (0.014)   &   0.268 (0.010)  &    0.256 (0.010)  &    0.264 (0.014)  &    0.290 (0.017)  &    0.230 (0.015)\\
{\bf Earth5: Solstice} & 0.462 (0.021)   &  0.346 (0.017)   &   0.268 (0.011)  &    0.257 (0.009)  &    0.270 (0.014)  &    0.298 (0.019)  &    0.230 (0.017)\\
{\bf Polar1: North}  & 0.575 (0.028)   &  0.464 (0.023)   &   0.373 (0.015)  &    0.368 (0.012)  &    0.388 (0.012)  &    0.425 (0.013)  &    0.355 (0.012)\\
{\bf Polar2: South}  & 0.590 (0.022)   &  0.475 (0.022)   &   0.383 (0.023)  &    0.372 (0.023)  &    0.382 (0.022)  &    0.414 (0.023)  &    0.351 (0.017)\\
 \hline
\end{tabular}

\medskip
The number in parentheses represents the root-mean-squared (RMS) time-variability of the apparent albedo in that waveband. For example, the Earth1 350~nm time series had a mean apparent albedo of 0.482 and an RMS variability of 0.016, or 3\%. 
\end{minipage}
\end{table*}

The time-averaged spectra for the EPOXI polar and equatorial observations are shown in Figure~\ref{average_albedo_spectrum}, and the apparent albedo values are listed in Table~\ref{apparent_albedos}.  The literature abounds with vaguely-defined ``reflectance'' measurements: as a result of differing definitions of reflectance used by different EPOXI team members, the albedos reported in \cite{Cowan_2009} were about $2/3$ of the correct value. (Note that this uniform offset had no impact on the color variations and analysis presented in that paper.)  

We present the corrected values of apparent albedo in Table~3 and Figure~3.  The major features of the time-averaged broadband albedo spectrum of Earth are: 1) a blue ramp shortward of 550~nm due to Rayleigh scattering, 2) a slight rise in albedo towards longer wavelengths due to continents, and 3) a steep dip in albedo at 950~nm due to water vapor absorption. Significantly, apart from a 20-30\% uniform offset, the polar and equatorial observations have indistinguishable albedo spectra.  

\section{Determining Principal Colors}
As in \cite{Cowan_2009} we assume no prior knowledge of the different surface types of the unresolved planet. Our data consist of 25 broadband spectra of Earth for each of two viewing geometries. For the equatorial observations \citep{Cowan_2009}, we found substantial variability in all wavebands (though the near-IR wavebands exhibited the most variability, leading to the dominant red eigencolor). The polar observations also show variability at all wavebands (Table~2), but as we argue below, the intrinsic cause of this variability is not necessarily the same surface types rotating in and out of view.

The multiband, time-resolved observations of Earth can be thought of as a locus of points occupying a 7-dimensional parameter space (one for each waveband). Principal component analysis (PCA) allows us to reduce the dimensionality of these data by defining orthonormal eigenvectors in color space (a.k.a.\ eigencolors). Quantitatively, the observed spectrum of Earth at some time $t$ can be recovered using the equation:
\begin{equation} \label{pca}
A^{*}(t, \lambda) = \langle A^{*}(t, \lambda)\rangle + \sum_{i=1}^{7} C_{i}(t) A_{i}(\lambda),
\end{equation}
where $\langle A^{*}(t, \lambda)\rangle$ is the time-averaged spectrum of Earth, $A_{i}(\lambda)$ are the seven orthonormal eigencolors, and $C_{i}(t)$ are the instantaneous projections of Earth's colors on the eigencolors. The terms in the sum are ranked by the time-variance in $C_{i}$, from largest to smallest. 

Insofar as the color variations are dominated by the first few terms of the sum, the locus does not occupy the full 7-dimensional color space, but a more restricted manifold. The dimensionality of the manifold is one fewer than the number of surface types rotating in and out of view.  E.g., a two-dimensional locus (a planar manifold) requires three surface types; a three-dimensional locus requires four surface types, etc. The general problem of estimating the pure surface spectra based on the morphology of such a locus of points is beyond the scope of this paper. It is a form of spectral unmixing, and is an area of active research in the remote sensing community \citep[e.g.,][]{LeMouelic_2009}.

In practice, there are two different ways to perform PCA, which may give quantitatively different results.  The analysis can be run using the covariance of the data, ${\rm Cov}(X,Y)= E[(X-E[X])(Y-E[Y])]$, where $E[X]$ is the expected value of $X$; or it can be run using the correlation of the data, ${\rm Corr}(X,Y)= {\rm Cov}(X,Y)/(\sigma_{X}\sigma_{Y})$, where $\sigma_{X}$ is the standard deviation of $X$. The correlation matrix is a standardized version of the covariance matrix; this is useful when the measured data do not all have the same units, since division by the standard deviation renders them unitless.  When the data are unitless to begin with, as is the case for our albedo measurements, running covariance-PCA is preferable \citep[e.g.,][]{Borgognone_2001}. In \cite{Cowan_2009} we used covariance-PCA, and we continue to do so here\footnote{For this paper, we use the Interactive Data Language (IDL) routine PCOMP (with the /COVARIANCE keyword set) to perform principal component analysis, while in \cite{Cowan_2009} we used the IDL routine SVDC, which performs singular value decomposition.}.

\subsection{Testing PCA on Simulated Data} \label{numerical_experiment}
Although PCA is a mathematically and numerically robust technique for analyzing patterns in data, interpreting its results can be ambiguous. In particular, we would like to verify to what extent there is a one-to-one correspondence between the eigencolors output by the PCA and real surfaces on Earth. To this end, we test the PCA routine on a suite of simulated data produced by the Virtual Planetary Laboratory's validated 3D spectral Earth model (details can be found in Robinson et al. 2011). The simulations used here were designed to closely mimic the Earth1 EPOXI observations taken in March 2008. 

We run five different versions of the VPL 3D Earth model: 1) Standard: this model is an excellent fit to the EPOXI Earth1 observations; the remaining models are identical, but in each case a single model element has been ``turned off'': 2) Cloud Free; 3) No Rayleigh Scattering; 4) Black Oceans; 5) Black Land.  We show the results of this experiment in Appendix I; here we simply state our conclusions: 

1) PCA successfully determines the dimensionality of the color variability and therefore the minimum number of different surface types contributing to color variations. In particular, $n$-dimensional variations require $n+1$ surface types (N.B. we count clouds as a surface type). 

2) Rayleigh scattering is important in determining the time-averaged broadband colors of Earth, but does not significantly affect its rotational color variability.

3) Cloud-free land surfaces, which are red, contribute a red eigencolor to the diurnal variability. The presence of relatively cloud-free land (deserts) near the equator explains why the rotational map of the red eigencolor \citep[Figure~10 in][]{Cowan_2009} successfully identified the major landforms and bodies of water on Earth.

4) Oceans are essentially a null surface, contributing neither to the broadband colors of Earth, nor to the time-variability of those colors, except insofar as the presence of oceans corresponds to a shortage of land.

5) In the absence of land, the variability is gray, due to large-scale inhomogeneities in cloud cover.

6) PCA necessarily outputs orthogonal eigencolors and a good deal of Earth's variability is due to clouds. Therefore, if the first eigencolor is red, then the second eigencolor may be blue even if there is no blue surface rotating in and out of view; this is an improvement on the interpretation of \cite{Cowan_2009}.

\subsection{Results of PCA for Polar Observations}
\begin{figure}[htb]
\includegraphics[width=84mm]{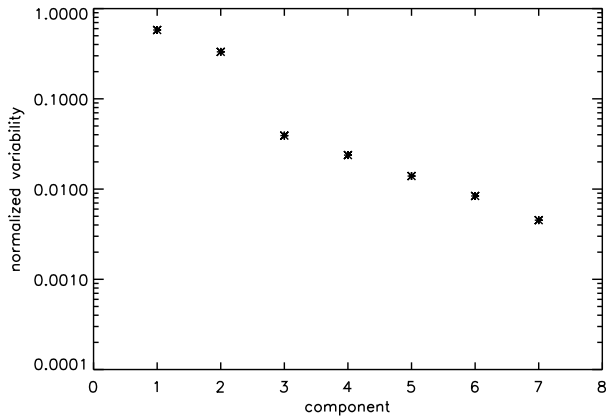}
\caption{{\bf North} Normalized variability in the 7 eigencolors of Earth's North polar regions, based on EPOXI observations taken on March 27, 2009. The color variations of Earth during these observations are well described as a combination of components 1 \& 2.}
\label{polar1_variability}
\end{figure}

\begin{figure}[htb]
\includegraphics[width=84mm]{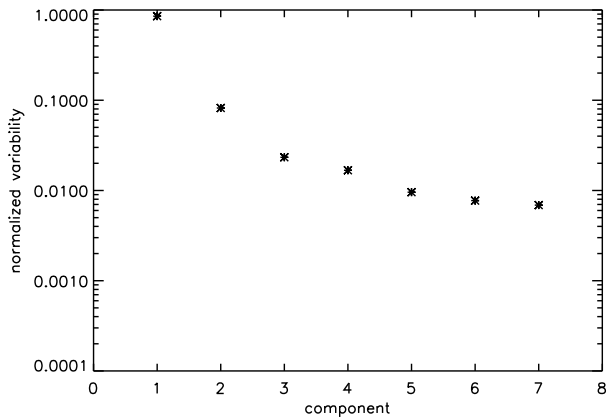}
\caption{{\bf South} Normalized variability in the 7 eigencolors of Earth's South polar regions, based on EPOXI observations taken on October 4, 2009.  The color variations of Earth during these observations are well described by component 1.}
\label{polar2_variability}
\end{figure}

In Figures~\ref{polar1_variability} \& \ref{polar2_variability} we show the eigenvalue spectra for time-variations in the 7 eigencolors identified by the PCA of the EPOXI polar observations. The eigenvalue for a given component is the projection of the data's variance onto that eigenvector; we plot here the square root of the eigenvalues, which is a measure of the RMS variability of the data projected onto an eigenvector. The variability has been normalized in the figures such that the sum of the variability for all seven components is unity. By definition, the low-order principal components have the largest variance. 

For the North observation, there are two eigencolors that dominate the color variations of Earth: the third eigencolor contributes only $\sim4$\% of the planet's color variability.  As in \cite{Cowan_2009}, this means that the colors of Earth populate a two-dimensional plane rather than filling the entire seven-dimensional color-space, and this requires at least three surface types. The southern observation, on the other hand, is dominated by a single eigencolor (the second eigencolor contributes to variability at the $<10$\% level).  This means that ---for the 24 hours of observations--- the colors of the planet populated a one-dimensional line in the seven-dimensional color volume, requiring only two surface types.

\begin{figure}[htb]
\includegraphics[width=84mm]{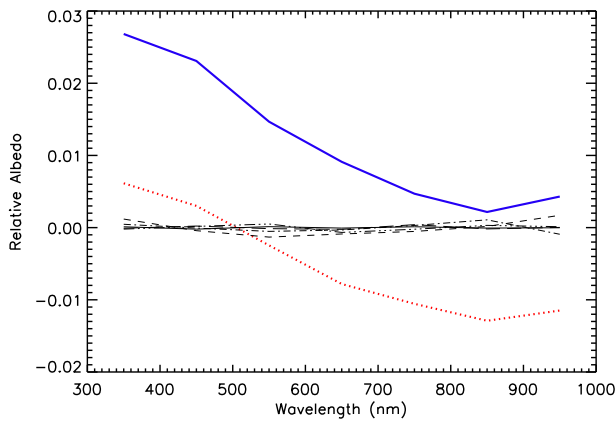}
\caption{{\bf North} Spectra for the eigencolors of northern Earth, as determined by PCA. The two dominant eigencolors are the bold solid and dotted lines.  The eigenspectra have been normalized by their eigenvalues, so the dominant components exhibit larger excursions from zero.}
\label{polar1_eigenspectra}
\end{figure}

\begin{figure}[htb]
\includegraphics[width=84mm]{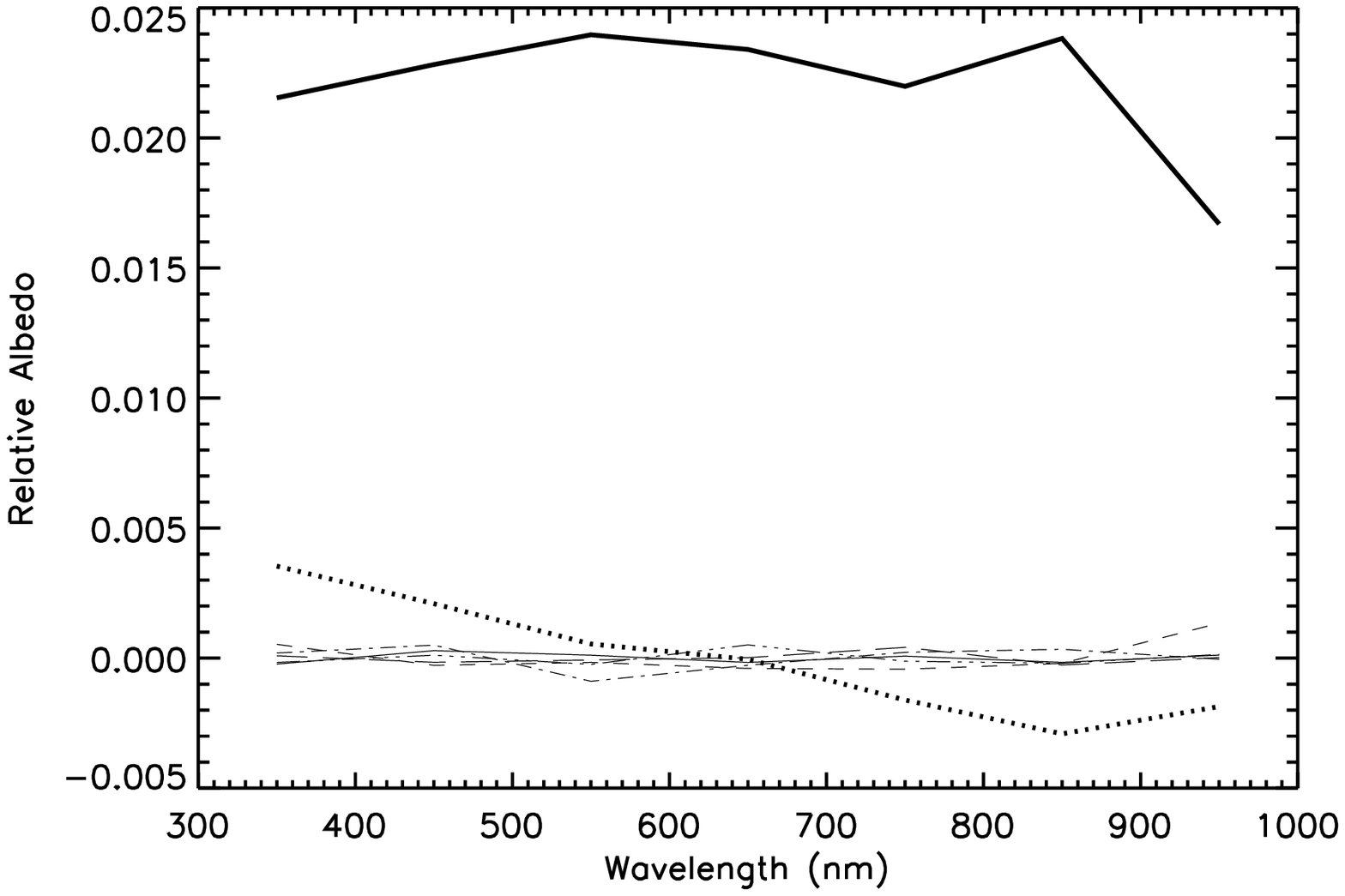}
\caption{{\bf South} Spectra for the eigencolors of southern Earth, as determined by PCA. The two dominant eigencolors are the bold solid and dotted lines. The eigenspectra have been normalized by their eigenvalues, so the dominant components exhibit larger excursions from zero.}
\label{polar2_eigenspectra}
\end{figure}

The eigencolors (the $A_{i}(\lambda)$ from Equation~\ref{pca}) are shown in Figures~\ref{polar1_eigenspectra} \& \ref{polar2_eigenspectra}. The raw eigencolors are ---by definition--- orthogonal and normalized ($\sum_{j=1}^{7}A_{i}^{2}(\lambda_{j})=1$), and this is how we presented them in \cite{Cowan_2009}. Here we have instead scaled the eigencolors by their associated eigenvalues ($\sum_{j=1}^{7}A_{i}^{2}(\lambda_{j})=v_{i}$, where $v_{i}$ is the eigenvalue of the $i$'th component), so the dominant components exhibit larger excursions from zero. 

\begin{figure}[htb]
\includegraphics[width=84mm]{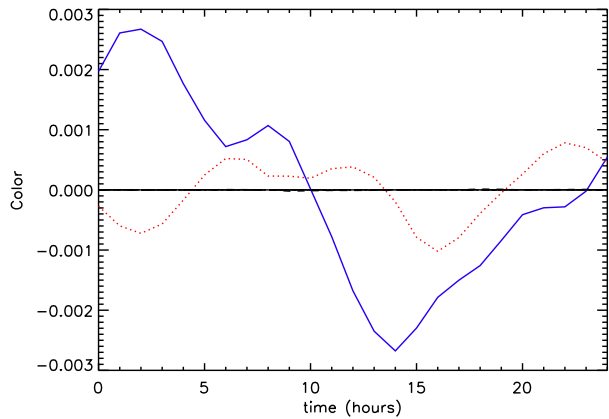}
\caption{{\bf North} Contributions of northern Earth's eigencolors, as determined by PCA, relative to the average Earth spectrum. The observations span a full rotation of the planet, starting and ending with the spacecraft directly above $152^{\circ}$~W longitude, the North Pacific Ocean.}
\label{polar1_eigenprojections}
\end{figure}

\begin{figure}[htb]
\includegraphics[width=84mm]{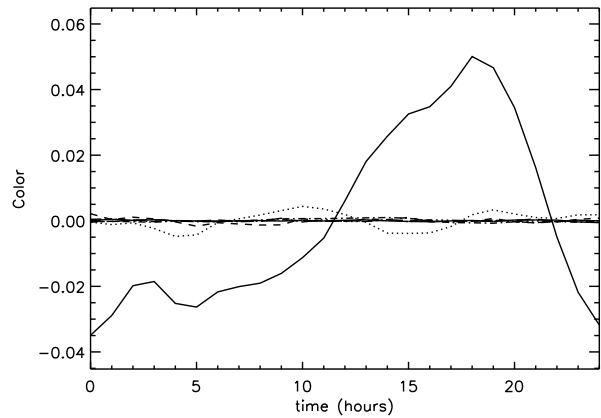}
\caption{{\bf South} Contributions of southern Earth's eigencolors, as determined by PCA, relative to the average Earth spectrum. The observations span a full rotation of the planet, starting and ending with the spacecraft directly above $59^{\circ}$~W longitude, in the South Atlantic Ocean.}
\label{polar2_eigenprojections}
\end{figure}

The eigenprojections ($C_{i}(t)$ from Equation~\ref{pca}) are shown in Figures~\ref{polar1_eigenprojections} \& \ref{polar2_eigenprojections}. The standard deviation of an eigenprojection corresponds to the variance or eigenvalue of that component. By definition, the low-order eigenprojections have the largest deviations from 0. Note that in \cite{Cowan_2009} we instead plotted the normalized eigenprojections ($\sum_{k=1}^{25}C_{i}(t_{k})^{2}=1$), which made it easier to compare the shapes of the eigenprojections but masked their relative importance.

The North polar observations are dominated by two eigencolors.  At first glance, the two eigencolors are identical, only offset in the vertical direction, but they are (by construction) orthogonal.  The more important of the two is blue, in that it is most non-zero at short wavelengths and nearly independent of what is going on at long wavelengths; the second eigencolor is red: it is most non-zero at long wavelengths and is largely insensitive to variability in blue wavebands. Based on the findings presented in \S~\ref{numerical_experiment}, we may infer that clouds and continents are rotating in and out of view as seen from this vantage point. Furthermore, cloud-related variability appears to be more important here than it was for the equatorial observations, which had a dominant red eigenvector followed by a blue, rather than vice-versa. 

The South polar observations are dominated by a single, gray eigencolor. Snow and clouds both have gray optical albedo spectra, so either may be contributing to the photometric variability. The absence of an important red eigencolor is due to the relative dearth of continents in the southern hemisphere. The second eigencolor is two orders of magnitude down in variance, or one order of magnitude in variability. It indicates that red and blue surfaces are trading places as the world turns ($A_{2}(\lambda)$ is positive at short wavelengths, negative at long wavelengths, and zero in between), but the forced orthogonality of the eigencolors makes this interpretation ambiguous. 

\section{Rotational Mapping} \label{viewing_geometry}
In this section we address how to infer the longitudinal color inhomogeneities of the unresolved planet based on time-resolved photometry. Note that this is in principle an independent question from that of identifying surface types on the planet (\S~3). One could try to infer the surface types on a planet without knowing or caring about their spatial distribution; or one could simply produce longitudinal color maps while remaining agnostic about what these tell us about surfaces (where ``surface'' here includes clouds). In practice, however, the two are intimately tied: a planet only exhibits rotational variability if it has a variegated surface \emph{and} substantial spatial inhomogeneities in the distribution of these surfaces.  

\subsection{Cloud Variability}
As in \cite{Cowan_2009}, we wish to estimate disk-integrated cloud variability, as this imposes a limit on the accuracy of any rotational maps we create\footnote{Note that diurnal cloud variability is not necessarily an obstacle for PCA, since that analysis is not predicated on a periodic signal.}. After 24 hours of rotation the same hemisphere of Earth should be facing the Deep Impact spacecraft, so the integrated brightness of the planet's surface should be nearly identical, provided one has accounted for the difference between the sidereal and solar day, as well as changes in the geocentric distance of the spacecraft and in the phase of the planet as seen from the spacecraft. Even after correcting for all known geometric effects, the observed fluxes at the start and end of our observing campaigns differ by  $\Delta A^{*}/\langle A^{*}\rangle =$3--6\% and 0.4--1\% for the North and South polar observing campaigns, respectively. We attribute this discrepancy to diurnal changes in cloud cover. 

Our 24-hour polar observation cloud variability of 4\% and 1\%, respectively is comparable to our estimate of cloud variability from previous EPOXI Earth observations 
\citep[2\% and 3\%,][]{Cowan_2009} and somewhat smaller than estimates from Earthshine observations. For example, \cite{Goode_2001} and \cite{Palle_2004} found day-to-day cloud variations of roughly 5\% and 10\%, respectively. Although we are still very much in the realm of small number statistics, it is conceivable that Earthshine observations over-estimate diurnal changes in cloud cover: under-estimating night-to-night calibration errors would lead to over-estimating day-to-day cloud variability.

Depending on the size of the telescope and cloud meteorology for a given planet, either photon counting or cloud variability can dominate the error budget for rotational inversion.  For the purposes of rotational mapping, we adopt effective Gaussian errors in the apparent albedo of $\sigma_{A^{*}}=0.01$, comparable to the actual day-to-day cloud variability on Earth.  

\subsection{Normalized Weight and the Dominant Latitude}
Using the formalism of \cite{Cowan_2009}, the visibility and illumination of a region on the planet at time $t$ are denoted by $V(\theta, \phi, t)$ and $I(\theta, \phi, t)$, respectively, where $\theta$ is the latitude and $\phi$ is the longitude on the planet's surface. $V$ is symmetric about the line-of-sight, is unity at the sub-observer point, drops as the cosine of the angle from the observer and is null on the far side of the planet from the observer; $I$ is symmetric about the star--planet line, is unity at the sub-stellar point, drops as the cosine of the angle from the star and is null on the night-side of the planet.

Following \cite{Fujii_2010} and \cite{Kawahara_2010}, we define the normalized weight,
\begin{equation}
W(\theta, \phi, t) = \frac{V(\theta, \phi, t) I(\theta, \phi, t)} {\oint V(\theta, \phi, t) I(\theta, \phi, t) d\Omega},
\end{equation}
which quantifies which regions of the planet are contributing the most to the observed light curve, under the assumption of diffuse (Lambertian) reflection. $W$ can be thought of as the smoothing kernel for the convolution between an albedo map, $A(\theta,\phi)$ and an observed lightcurve, $A^{*}(t)$.  

To first order, the character of a lightcurve can be understood in terms of the shape and location of the weight function. The normalization (denominator) of the weight is a simple function of phase:
\begin{equation}
\oint{V(\theta,\phi,t) I(\theta,\phi,t) d\Omega} = \frac{2}{3} \left[\sin\alpha + (\pi-\alpha)\cos\alpha \right].
\end{equation}

As the planet rotates, $W$ sweeps from East to West. The width (in longitude) of the weight determines the longitudinal resolution achievable by inverting diurnal light curves: a broad $W$ at full phase leads to a coarser map than the slender $W$ of a crescent phase. This of course neglects the practical issues of inner working angles and photon counting noise. 

The peak of $W$ lies half-way between the sub-stellar and sub-observer points and corresponds to the location of the glint spot. The latitude of the glint spot may change throughout an orbit: the sub-observer latitude is fixed in the absence of precession, but the sub-solar latitude exhibits seasonal changes for non-zero obliquity.

The peak of the weight is the area of the planet that contributes the most to the observed disc-integrated light curve. E.g., a polar sub-observer latitude and an equatorial sub-stellar latitude would yield a weight with a maximum at mid-latitudes. In detail, $W$ is also tempered by the usual $\sin\theta$ dependence of $d\Omega$ (ie: there is more area near the equator than near the poles). The dominant latitude (the latitude where the most photons would originate from in the case of a uniform Lambert sphere) is therefore not simply the peak of $W$, but is rather the average $\theta$, weighted by $W$:
\begin{equation}
\theta_{\rm dom}(t) = \oint{W(\theta, \phi, t) \theta d\Omega}= \frac{\oint{V I \theta d\Omega}}{\oint{V I d\Omega}}.
\end{equation}

In Table~1 we list the dominant latitude for the four EPOXI observations. Significantly, the dominant latitude is temperate for the ``polar'' observations, despite the exotic viewing geometry.  This simple argument explains why the time-averaged colors of the polar EPOXI Earth observations are so familiar: most of the photons will not originate from the snowy and icy regions of Earth. 

That being said, the mid-latitudes probed by the polar observations are significantly more cloudy than the tropics \citep[yearly mean cloud cover in the tropics is 25--50\%, while at 45$^{\circ}$~S cloud cover is 75--100\%; see Figure~6a of][]{Palle_2008}. As shown in Appendix~I, clouds contribute a uniform (gray) increase in albedo of 20--50\% between the cloud-free and standard VPL models, so a latitudinal difference in cloud cover is a natural explanation for the observed difference in albedo. 

Furthermore, polar snow and ice necessarily contribute more to the polar than the equatorial EPOXI observations.  For example, the Earth1 and Polar1 EPOXI observations were both obtained at the same time of year (March 2008 and 2009, respectively), so we may meaningfully ask how the different viewing geometries affect the contribution from snowy regions. If the global-mean snowline for March lies at $55^{\circ}$N, we find that 2\% of the weight is in snow-covered regions for the equatorial observation, while this fraction is 16\% for the polar observation. In general, for global mean snow lines between 50--60$^{\circ}$N, the snow-covered regions contribute 7--9 times more to the polar observation than to the equatorial observation. But such an argument is unlikely to work for the Polar2 observation, which probes the relatively land-free southern oceans, and the time-averaged Polar1 and Polar2 broadband colors are indistinguishable.  We therefore believe the increased weight of clouds to be the main source of the 20--30\% greater absolute value of $A^{*}$ in the polar observations compared to the equatorial observations.

\begin{figure}[htb]
\includegraphics[width=84mm]{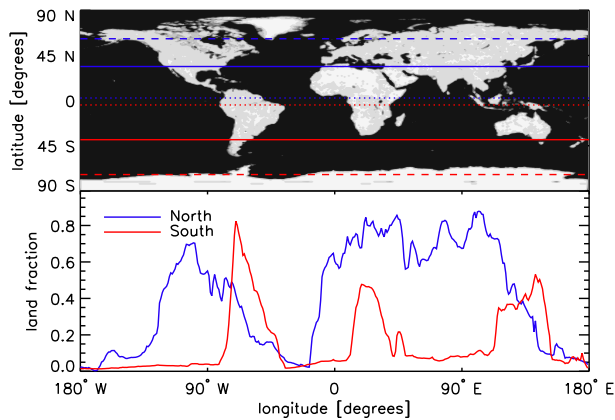}
\caption{\emph{Top:} Land coverage map for modern Earth.  The colored lines indicate important latitudes for the EPOXI North (blue) and South (red) polar observations. The dotted lines show the sub-solar latitudes; the dashed lines show the sub-observer latitudes; the solid lines show the dominant latitudes, which are expected to contribute the most to the lightcurves. \emph{Bottom:} The longitudinal land coverage profiles for the EPOXI Polar observations.}
\label{polar_earth_map}
\end{figure}

In the top panel of Figure~\ref{polar_earth_map}, we show a map of land coverage on Earth and indicate the sub-solar, sub-observer, and dominant latitudes for the North and South polar observations. The bottom panel of the figure shows the longitudinal land fraction profiles for the two polar observations, obtained by integrating the 2-D map by the weight function dictated by viewing geometry. It indicates the location of the major landforms probed by the polar observations and can be compared to the longitudinal profiles of eigencolors presented in the following section.    

\subsection{Lightcurve Inversion}
The flux ratio primarily depends on the planet's orbital phase, the observer--planet--star angle, and the ratio $(R_{p}/a)^2$. We define the apparent albedo, $A^{*}$, as the ratio of the flux from the planet divided by the flux we would expect \emph{at the same phase} for a perfectly reflecting Lambert sphere \citep[see also][]{Qiu_2003}. This amounts to the average albedo of the planet, weighted by $W$: 
\begin{equation}
A^{*}(t, \lambda) = \oint{W(\theta, \phi, t) A(\theta, \phi, \lambda) d\Omega} = \frac{\oint V I A d\Omega}{\oint V I d\Omega}.
\end{equation}
A uniform planet would have an apparent albedo that is constant over a planetary rotation in the $P_{rot} \ll P_{orb}$ limit; a uniform Lambert sphere would further have a constant apparent albedo during the entire orbit. For non-transiting exoplanets, the planetary radius may be unknown, in which case $A^{*}$ can only be determined to within a factor of $R_{p}^{2}$, with a lower limit on the radius obtained by setting $A^{*}=1$.

Lightcurve inversion means inferring $A(\theta,\phi, \lambda)$ from $A^{*}(t, \lambda)$. If observations only span a single rotation, or if a planet has zero obliquity, one can only constrain the longitudinal variations in albedo.

The visibility, $V(\theta,\phi,t)$, and illumination, $I(\theta,\phi,t)$, can be expressed compactly in terms of the locations of the sub-observer and sub-stellar points:
\begin{equation}
\begin{array}{ll} V(\theta,\phi,t) = &\max[\sin\theta \sin\theta_{\rm obs} \cos(\phi - \phi_{\rm obs}) \\
& + \cos\theta \cos\theta_{\rm obs}, 0]\\
I(\theta,\phi,t) = & \max[\sin\theta \sin\theta_{\rm star} \cos(\phi-\phi_{\rm star}) \\
& + \cos\theta \cos\theta_{\rm star}, 0],
\end{array}
\end{equation}
where $\phi_{\rm obs}(t) = \phi_{\rm obs}(0) - \omega_{\rm rot}t$ is the sub-observer longitude, $\theta_{\rm obs}$ is the constant sub-observer latitude, $\phi_{\rm star}(t)$ and $\theta_{\rm star}(t) = \arccos[\cos(\xi_{0}+\omega_{\rm orb}t-\xi_{\rm obl})\sin\theta_{\rm obl}]$ are the sub-stellar longitude and latitude; $\omega_{\rm rot}$ and $\omega_{\rm orb}$ are the rotational and orbital angular velocities of the planet, $\xi_{0}$ is the initial orbital position of the planet, $\xi_{\rm obl}$ is the orbital location of northern summer solstice, and $\theta_{\rm obl}$ is the planet's obliquity. It is non-trivial to compute $\phi_{\rm star}(t)$ over a sizable fraction of an orbit, requiring a numerical integration or use of the equation of time. 

For the current application, however, the planet's rotation period is much shorter than its orbital period, so it is sufficient to assume that $\theta_{\rm star}$ is constant and $\phi_{\rm star}(t)$ advances linearly at one revolution per solar day (as opposed to the sidereal day used in computing $\phi_{\rm obs}(t)$). We use Horizons\footnote{http://ssd.jpl.nasa.gov/horizons.cgi} to compute the relative positions of the Deep Impact spacecraft, Earth and the Sun at the start of the various EPOXI campaigns. 

Both of the EPOXI polar observations were obtained with a viewing geometry very close to quadrature\footnote{For small obliquities, polar observations imply that the planet is in a nearly face-on orbit, and therefore permanently at quadrature. But there is no reason to assume low obliquity for terrestrial planets, so polar observations need not be made at quadrature.}. The weight function therefore has a width of 90$^{\circ}$ in longitude, indicating that we would need a model with  8 longitudinal slices of uniform albedo to achieve Nyquist sampling of the rotational lightcurve, in the absence of specular reflection \citep[for more discussion on slice vs sinusoidal longitudinal profiles see][]{Cowan_2008, Cowan_2009}.  An 8-slice model with variable phase offset (prime meridian) would have 9 model parameters; we instead use sinusoidal maps with terms up to fourth order, which also have 9 model parameters but which converge better \citep{Cowan_2008}. The best-fit reduced $\chi^{2}$ of these models are somewhat lower than unity because the $\sigma_{A^{*}}=0.01$ ``error bars'' that we use are much larger than the point-to-point scatter in the lightcurves.

We estimate uncertainties in our longitudinal eigencolor maps with a Monte Carlo test. Using our adopted photometric error of $\sigma_{A^{*}}=0.01$, we generate 10,000 statistically equivalent instances of the observed lightcurves assuming Gaussian, uncorrelated errors. We run the same PCA and lightcurve inversion on each of these mock data sets and take the standard deviation in the resulting maps to be the uncertainty in our fiducial maps.
 
\begin{figure}[htb]
\includegraphics[width=84mm]{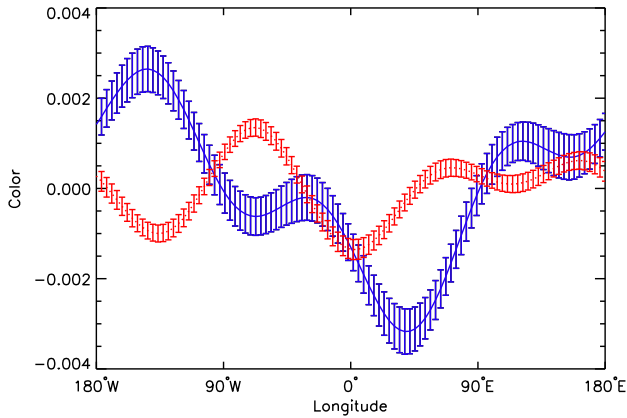}
\caption{{\bf North:} Longitudinal profiles of the two dominant eigencolors of Earth based on the lightcurves in Figure~\ref{polar1_lightcurves}, the eigencolors shown in Figure~\ref{polar1_eigenspectra}, and the known phase and rotational period of Earth.}
\label{polar1_eigen_maps}
\end{figure}

\begin{figure}[htb]
\includegraphics[width=84mm]{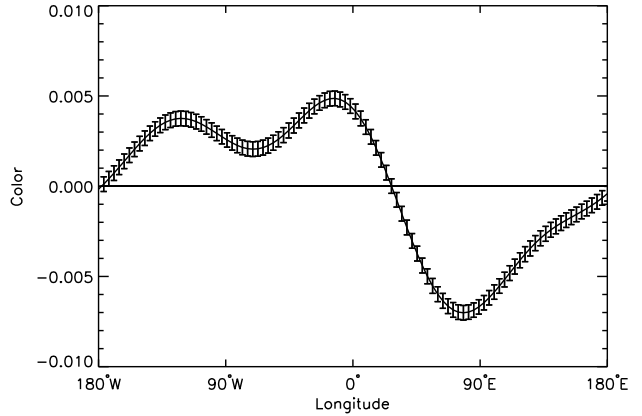}
\caption{{\bf South:} Longitudinal profiles of the dominant eigencolor of Earth based on the lightcurves in Figure~\ref{polar2_lightcurves}, the eigencolors shown in Figure~\ref{polar2_eigenspectra}, and the known phase and rotational period of Earth.}
\label{polar2_eigen_maps}
\end{figure}

Note that the rotational inversion may be performed directly on the lightcurves shown in Figures 1 \& 2, and independently of the PCA described in \S~3, yielding albedo maps of Earth in various wavebands. Instead, we combine PCA and lightcurve inversion as we did in \cite{Cowan_2009} and produce longitudinal maps of the dominant eigencolors, shown in Figures~\ref{polar1_eigen_maps} \& \ref{polar2_eigen_maps}. Based on the numerical experiments of \S~3.1, we expect both dominant eigencolors in Figure~\ref{polar1_eigen_maps} to be tracking clouds and snow-covered land, while the red eigencolor is also sensitive to cloud-free land. Since clouds are more prevalent at these latitudes, the red eigencolor does not faithfully locate the major landforms, as it did in \cite{Cowan_2009}. The southern polar observation (Figure~\ref{polar2_eigen_maps}) shows a broad maximum in the gray eigencolor at a longitude of 90$^{\circ}$~W, roughly corresponding to the location of Patagonia and the Antarctic Peninsula. Since snow-covered land is essentially indistinguishable from clouds at these poor spectral resolutions, we must remain agnostic about the source of this variability. 

In general, the cloudy mid-latitudes keeps the eigencolor maps from faithfully identifying major landforms (eg, compare Figures~\ref{polar1_eigen_maps} and \ref{polar2_eigen_maps} to Figure~\ref{polar_earth_map}).

\section{Albedo Model of Snowball Earth}
Since the polar regions of Earth are largely covered in snow and ice, it is worth asking if one might confuse a habitable planet like Earth with a snowball planet \citep[ie: one caught in the cold branch of a snow-albedo positive feedback loop.  See, for example, ][and references therein]{Tajika_2008}. In this section we describe a toy model for the reflectance of such a snowball planet, and compare the resulting photometry to the EPOXI polar observations. Note that \cite{Vasquez_2006} presented \emph{bolometric} (white light) diurnal lightcurves for a model snowball Earth, but these are not useful for the current comparison.

\subsection{Geography}
\begin{figure}[htb]
\includegraphics[width=84mm]{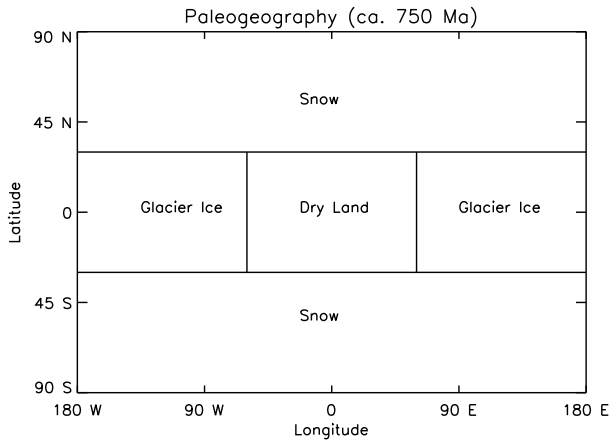}
\caption{The layout of surface types in our Snowball Earth model. Regions within 60$^{\circ}$  of the poles ---both continents and oceans--- are covered in snow. The tropics are dry (bare land or glacier ice) due to negative net precipitation.}
\label{continents_map}
\end{figure}

The geography of our snowball Earth model is shown in Figure~\ref{continents_map}. We use the same idealized paleogeography for the Sturtian glaciation ($\sim$750~Mya) as \cite{Pierrehumbert_2005}.  Paleomagnetism only constrains the magnetic \emph{latitude} of continents, however, and we are at liberty to choose any longitudinal distribution. The diurnal variability of the planet is determined solely by its longitudinal geography, however. If the continents are spread out uniformly in longitude, for example, the planet would not exhibit any rotational variability in the absence of heterogeneous cloud cover. We instead adopt the opposite limit of a single mega-continent. This will tend to exaggerate the amplitude of the diurnal variations in apparent albedo, but the changes in color should be robust. 

Assuming that sea-level was not grossly different from today, and that only trace amounts of continent-formation has occurred in the intervening 750 Myrs, continents should cover $\sim 25$\% of the planet, as today. At first sight, assuming constant water levels during a global glaciation seems inconsistent.  However, it is only icecaps (ice on land) that significantly change water levels, while the geological evidence for snowball Earth episodes instead require that the \emph{oceans} be frozen at the equator.  In fact, this criterion effectively shuts down the planet's hydrological cycle, so the polar ice caps are not very different from the present day. In any case, the high latitudes of a snowball Earth should be covered in snow, regardless of whether the underlying regions are continent or ocean.  Therefore, the precise fraction of land vs ocean does not directly impact the observed diurnal variability. 

For a snowball planet, we assume that both oceans and continents are covered in snow at latitudes greater than 30$^{\circ}$.  Closer to the equator, we assume that continents are bare, dry land due to the very low precipitation; while oceans are covered in blue glacier ice that flows towards the equator \citep{Goodman_2003}.    
  
\subsection{Albedo Spectra}
Climate models of Snowball Earth are concerned with albedo only insofar as it modulates the energy budget of the planet.  That is to say, they care about the Bond albedo, $A_{\rm B}$, the fraction of incident energy that is reflected back into space. We, on the other hand, are concerned with the appearance of the planet as seen from the outside, $A^{*}(t,\lambda)$. It is beyond the scope of this paper to make a detailed snowball earth spectral model. We therefore assume diffuse (a.k.a. Lambert) reflection and use a wavelength-dependent albedo, $A(\lambda)$, that is simply a function of location rather than a bidirectional reflectance distribution function. A hard Snowball Earth will not have appreciable expanses of liquid water to contribute to glint. That being said, other elements, notably clouds, are not strictly Lambertian \citep[eg,][]{Robinson_2010}.      

For a cloud-free and airless model planet, the albedo spectrum at each point on the planet is simply determined by the albedo spectrum of the surface type at that location.  Reflection from clouds and Rayleigh scattering from air molecules complicates this picture, however. We treat the albedo from these semi-transparent media as follows:
\begin{equation}\label{combine_albedo}
A(\lambda) = 1 - (1-A_{\rm Ra}(\lambda))(1-A_{\rm cl}(\lambda))(1-A_{\rm surf}(\lambda)),
\end{equation}    
where $A_{\rm Ra}$ is the effective albedo due to Rayleigh scattering, $A_{\rm cl}$ is the albedo due to clouds, and $A_{\rm surf}$ is the albedo of the surface type at that point on the planet. This simple expression captures the essential behavior of clouds and Rayleigh scattering: they always increase the effective albedo of a region, but the effect is most pronounced for a dark underlying surface.

\begin{figure}[htb]
\includegraphics[width=84mm]{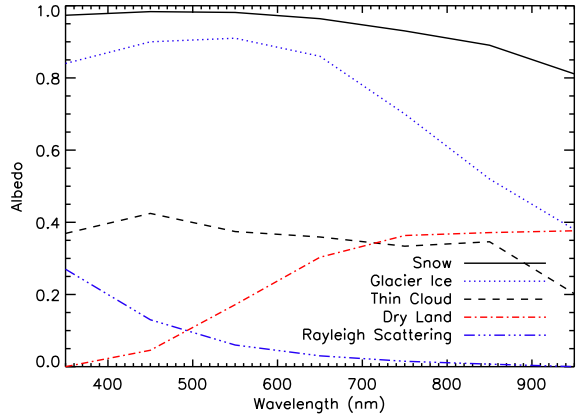}
\caption{The broadband optical albedo spectra for the components of our snowball Earth model. The Bond albedos of the surfaces are estimated using a Solar spectrum and integrating to 5~$\mu$m (snow: 0.8; glacier ice: 0.6; thin cloud: 0.3, dry land: 0.3, Rayleigh scattering: $<0.1$).}
\label{surface_albedo_spectra}
\end{figure}

Our model has 5 elements, each with a distinctive albedo spectrum, shown in Figure~\ref{surface_albedo_spectra}. We use spectra for dry land and snow from Robinson et al. (2011). The snow albedo spectrum we use is for medium grained snow, while the cold, dry climate of a snowball Earth would create small-grained snow, as seen in Antarctica \citep{Hudson_2006}. There is no perceptible difference in the broadband albedo spectra of these two kinds of snow at optical wavelengths, however. We use the empirical albedo spectrum for blue glacier ice from \cite{Warren_2002}. To mimic the thin clouds expected on a frozen planet with reduced hydrological activity, we take a generic cloud spectrum (Robinson 2009, priv. comm.) and divide the albedo by 2. (Using thicker clouds increases the Bond albedo of our model planet but does not significantly change the color variability.) We distribute the clouds on the planet using a snapshot of cloud maps from a snowball Earth general circulation model \citep{Abbot_2010}. Note that this model was run using the same idealized geography shown in Figure~12 and thus offers a good estimate of the spatial ---and in particular longitudinal--- variations in cloud cover. We estimate the disc-integrated effect of Rayleigh scattering by comparing the Standard and Rayleigh-Scattering-Free VPL models (described in Appendix~I) and using Equation~\ref{combine_albedo}.   

\begin{figure}[htb]
\includegraphics[width=84mm]{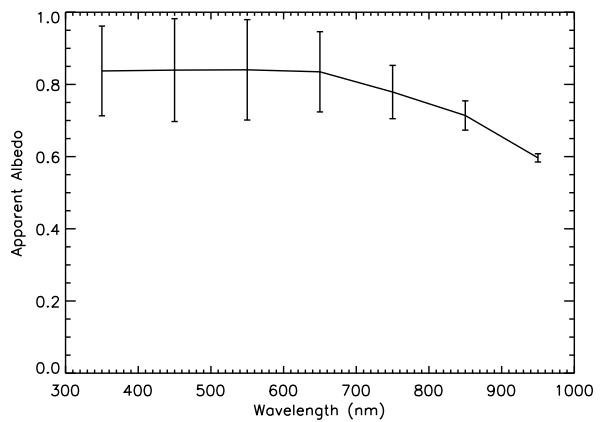}
\caption{Time-averaged albedo spectrum of Snowball Earth. The vertical bars show the RMS variability in each band. Relative peak-to-trough variability ranges from 42\% (450~nm) to 6\% (950~nm).}
\label{time_averaged_spectrum}
\end{figure}

Our snowball Earth model has a time-averaged Bond albedo of 70\%, which is self-consistent with the snow-albedo feedback\footnote{Despite its name, a snowball planet's albedo does not equal that of snow (80\%).  The cold, dry atmosphere keeps the tropical land bare ($A_{B}=30\%$) and exposes glacier ice ($A_{B}=60\%$) near the equator.}: \cite{Wetherald_1975} used a 70\% albedo to induce their ``White Earth'' solution in a 2-D model. \cite{Pierrehumbert_2005} ran GCMs of a hard snowball Earth and maintained the snowball state with albedos of 60--67\%. \cite{Chandler_2000}, on the other hand, ran snowball Earth global climate models (GCMs) with bolometric albedos of 20--40\%; \cite{Vasquez_2006} have an average Bond albedo of approximately 50\% for the frozen Earth. 

\subsection{Time Variability}
For directly-imaged exoplanets, the albedo cannot be determined independently of the planet's radius: photometry of reflected light will constrain the quantity $A R_{p}^{2}$. Therefore, while the general agreement in $A_{B}$ between our toy model and self-consistent simulations is encouraging, one cannot in general use the absolute albedo of an exoplanet as a diagnostic. Multiband observations will tell us about the colors of the planet, however, and with sufficiently high cadence observations it will be possible to measure the variations in apparent albedo due to the planet's rotation.

We compute the variations in apparent albedo for an equatorial observer and a planet at quadrature. The time-averaged spectrum of our snowball Earth model (Figure~\ref{time_averaged_spectrum}) is completely different from that of the modern Earth, regardless of viewing geometry (Figure~\ref{average_albedo_spectrum}). The flat spectrum shortward of 650~nm is due to snow and glacier ice, which are so reflective at these wavelengths as to make Rayleigh scattering imperceptible. The drop in albedo at longer wavelengths is also driven by snow and glacier ice. 

For our snowball model, the shortest wavelengths exhibit the most variability, as shown by the vertical bars in Figure~\ref{time_averaged_spectrum}. This is because bare land and glacier ice exhibit the largest contrast at short wavelengths, while at longer wavelengths they both have near-IR albedos of $\sim$40\%. This is in stark contrast to the case for the modern Earth, which exhibits variability at all wavebands.  We conclude that  ---given high-quality photometry--- the modern-day Earth could not be mistaken for a snowball planet, regardless of the viewing geometry. 

\section{Summary}
We presented time-resolved, disc-integrated observations of Earth's polar regions from the Deep Impact spacecraft as part of the EPOXI Mission of Opportunity. These complement the equatorial views presented in \cite{Cowan_2009}. We found that both of the polar observations have broadband colors similar to the equatorial Earth, but with uniformly higher albedos. We explained this in terms of the 2-D weight function for disc-integrated observations of Earth, which was most sensitive to the tropics for the equatorial observations, and most sensitive to mid-latitudes for the polar observations.

We performed PCA on a suite of simulated rotational multi-band lightcurves from NASA's Virtual Planetary Laboratory 3D Earth model. We found that PCA correctly indicates the number of different surfaces rotating in and out of view. We found that while the red eigencolor consistently tracks cloud-free land, a blue eigencolor only tracks oceans when clouds are entirely absent from the simulation. In the general (cloudy) case, a blue eigencolor is simply tracking cloud inhomogeneities. Gray eigencolors, when they are present, track large cloud patterns and/or snow-covered land. 

We also performed PCA on the EPOXI polar observations. Comparing these eigencolors to known surface types on Earth, we establish that the variability seen in the North EPOXI Polar observation is due to clouds, continents and oceans rotating in and out of view; the lack of large cloud-free land (ie: deserts) at the latitudes probed by these observations keep us from being able to faithfully extract the positions of major northern landforms. The South polar observation, on the other hand, was characterized by gray variability due to a large cloud pattern in the south oceans.

Lastly, we constructed a simple reflectance model for a snowball Earth, and found that both the time-averaged broadband colors and diurnal color variations of a Snowball Earth (gray snow $+$ near-IR roll-off; variability at blue wavelengths) would be distinguishable from the modern-day Earth  (near-UV Rayleigh ramp $+$ gray clouds; variability at all wavelengths), regardless of viewing angle.

\section{Discussion}
We listed the possible constraints one could get from time-resolved photometry in Section~1, here we briefly review the three forms of photometric characterization used in this paper in the context of measurements from planned space telescopes.

We have considered time-resolved, multi-band optical photometry, which could be obtained with a space-based high-contrast imaging mission.  On its own, such a telescope could discover nearby exoplanets, determine their approximate orbits, and characterize their time-averaged colors and time-variability of colors.

The \emph{time-averaged colors} of a planet will be the most accessible observable. We have shown in this paper that the broadband colors would be sufficient to distinguish between the modern day Earth and a snowball Earth. On modern-day Earth, polar ice and snow don't contribute significantly to the time-averaged albedo of Earth ---even with polar viewing geometries--- because of the glancing angle of sunlight at those latitudes. The coldest regions of a planet receive the least sunlight, and therefore contribute correspondingly little to the disk-integrated properties of the planet. 

The \emph{time-variability of the colors} are harder to obtain, requiring shorter integration times and therefore a larger telescope, all other things being equal. On the other hand, variability measurements are more robust to contamination from exo-zodiacal light. We have shown that modern Earth ---regardless of viewing angle--- exhibits photometric variability at all wavelengths (RMS variability within a factor of 2 for all wavebands), while Snowball Earth varies 7 times more at short wavelengths than at long wavelengths. A more subtle analysis of the time variability may even allow us to distinguish between the equatorial and polar viewing geometries of Earth,  because clouds play a larger role at mid-latitudes.  From an equatorial vantage point, the dominant eigencolor is red, followed by blue; for the polar geometries, the ordering of the red and blue eigencolors is flipped, or there is a single dominant grey eigencolor.

If the planet's radius can be estimated, then its albedo can be put on an absolute scale and one can estimate its Bond albedo.  Transiting planets have very well characterized radii, but nearby earth analogs will almost certainly \emph{not} be transiting. Instead, a radius estimate will require an additional large space mission: either an infrared high-contrast imaging telescope, or a space based astrometry mission.  In the first case, thermal and reflected photometry can be combined to estimate the planet's radius.  If thermal photometry is obtained at a variety of phases, then the efficiency of heat transport to the planet's night-side may be estimated \citep{Cowan_2011b}, and the systematic uncertainty will be $\sim2$\% due to the unkown efficiency of latitudinal heat transport \citep{Cowan_2011}. The uncertainty in the radius will therefore likely be dominated by the ---known--- uncertainties in thermal and reflected photometry.   

In the second case, the star's astrometric wobble provides a mass measurement for the planet; by assuming a planetary density, one can estimate the planet's radius.  The dominant source of uncertainty here is the planet's composition: given a mass, a planet's radius may vary by 50\% \citep[eg,][]{Charbonneau_2009, Batalha_2011}, leading to absolute albedo estimates only valid to within a factor of 2.  Transiting planet surveys will likely reduce these systematic uncertainties by providing an empirical mass-radius relation for planets across a wide range of masses. It is not clear to what extent the \emph{Kepler} mission \citep[eg,][]{Borucki_2011} will help define the mass-radius relation: although the vast majority of \emph{Kepler} candidates are likely to be \emph{bona fide} planets \citep{Morton_2011}, most will not have mass estimates. The smaller, and better characterized, radius uncertainty would therefore most likely come from combining optical and infrared photometry, rather than from a mass measurement.

The \emph{Bond albedo} would allow us to better distinguish between the equatorial ($A_{B}\approx 0.3$) and polar ($A_{B}\approx 0.4$) EPOXI observations, or between a snowball planet ($A_{B}\approx 0.7$) and a temperate one. In general, this quantity would be very useful in determining a planet's energy budget and would go a long way towards constraining its habitability.

\acknowledgments
This work was supported by the NASA Discovery Program. We thank D.S.~Abbot and R.T.~Pierrehumbert for providing us with cloud maps of snowball Earth. N.B.C.\ acknowledges many useful discussions with S.G.~Warren about snowball Earth, and thanks W.~Sullivan for encouraging him to complete his astrobiology research rotation. E.A.\ is supported by a National Science Foundation Career Grant.

\section*{Appendix I: PCA of Simulated VPL Data}
For completeness, we begin by re-running the PCA on the actual Earth1 data obtained by the Deep Impact spacecraft.  This endeavor is not redundant, since there are many differences between our current analysis and that of \cite{Cowan_2009}: 1) here we use a different solar spectrum in computing reflectance (the change is most noticeable in the 950~nm waveband); 2) we are now using rigorously-defined apparent albedo; 3) we run the analysis individually on the Earth1, rather than on both equatorial observations simultaneously; and 4) we do not apply the cloud-variability uncertainties when running the PCA. 

\begin{figure*}[htb]
\includegraphics[width=84mm]{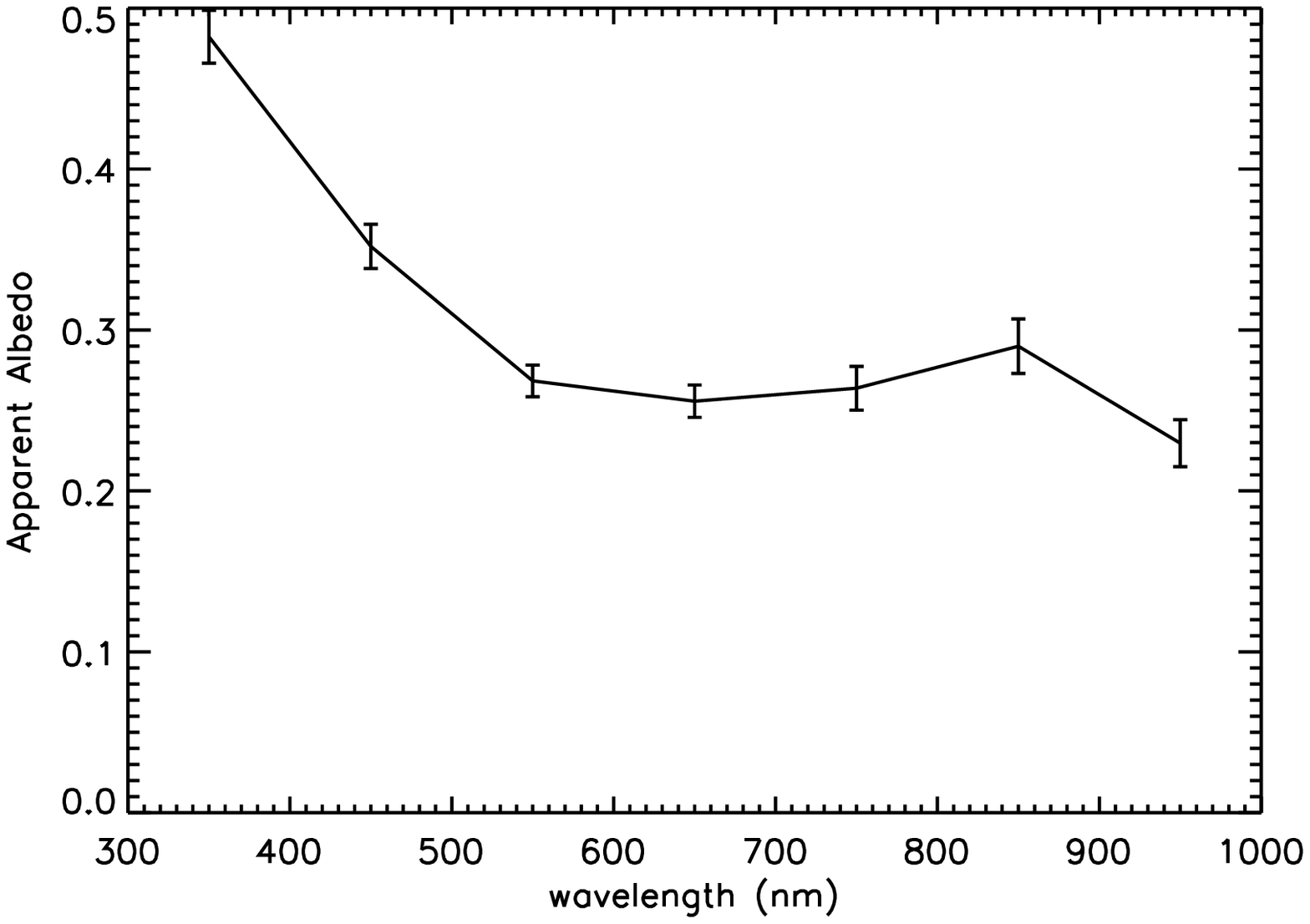}
\includegraphics[width=84mm]{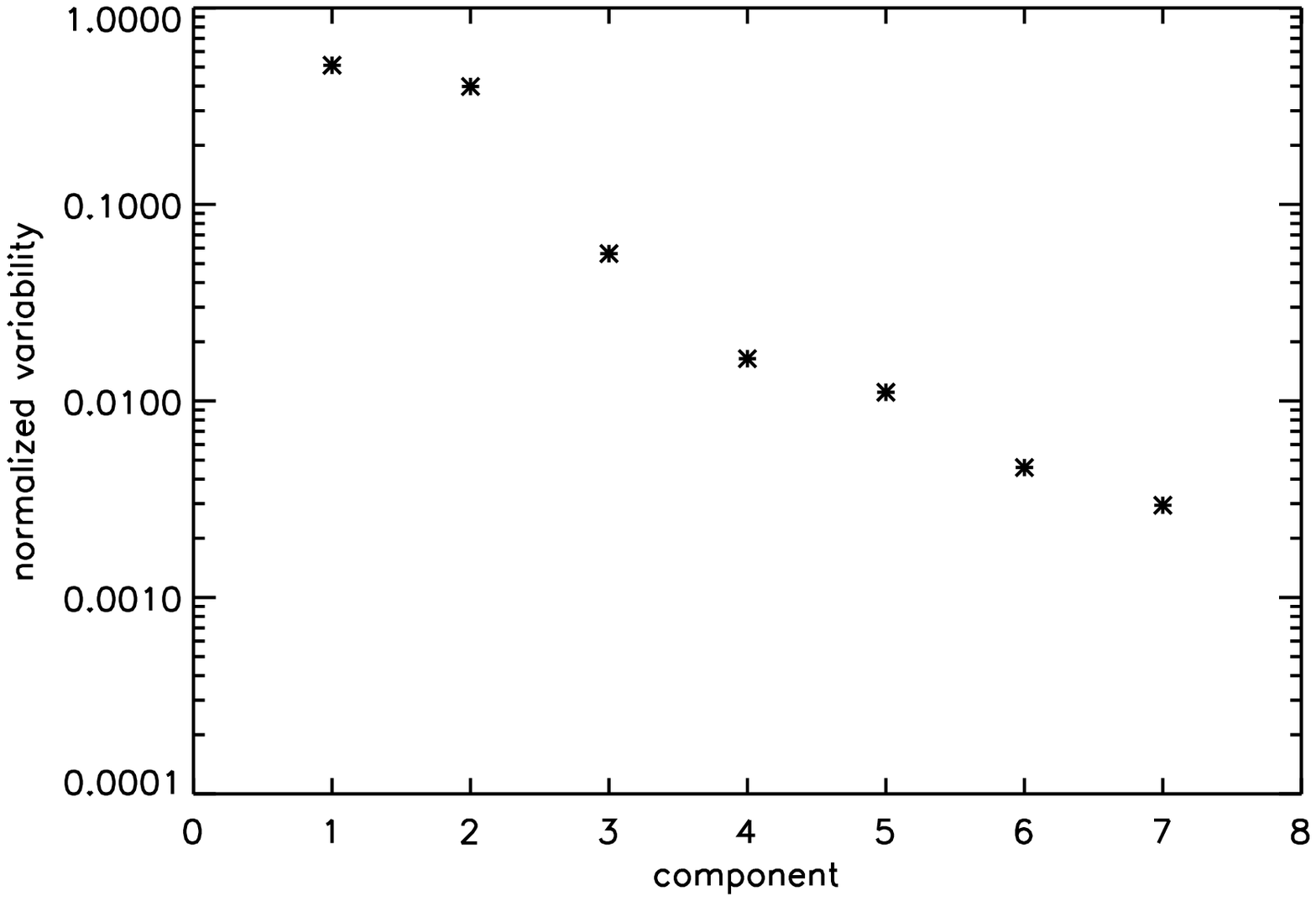}
\includegraphics[width=84mm]{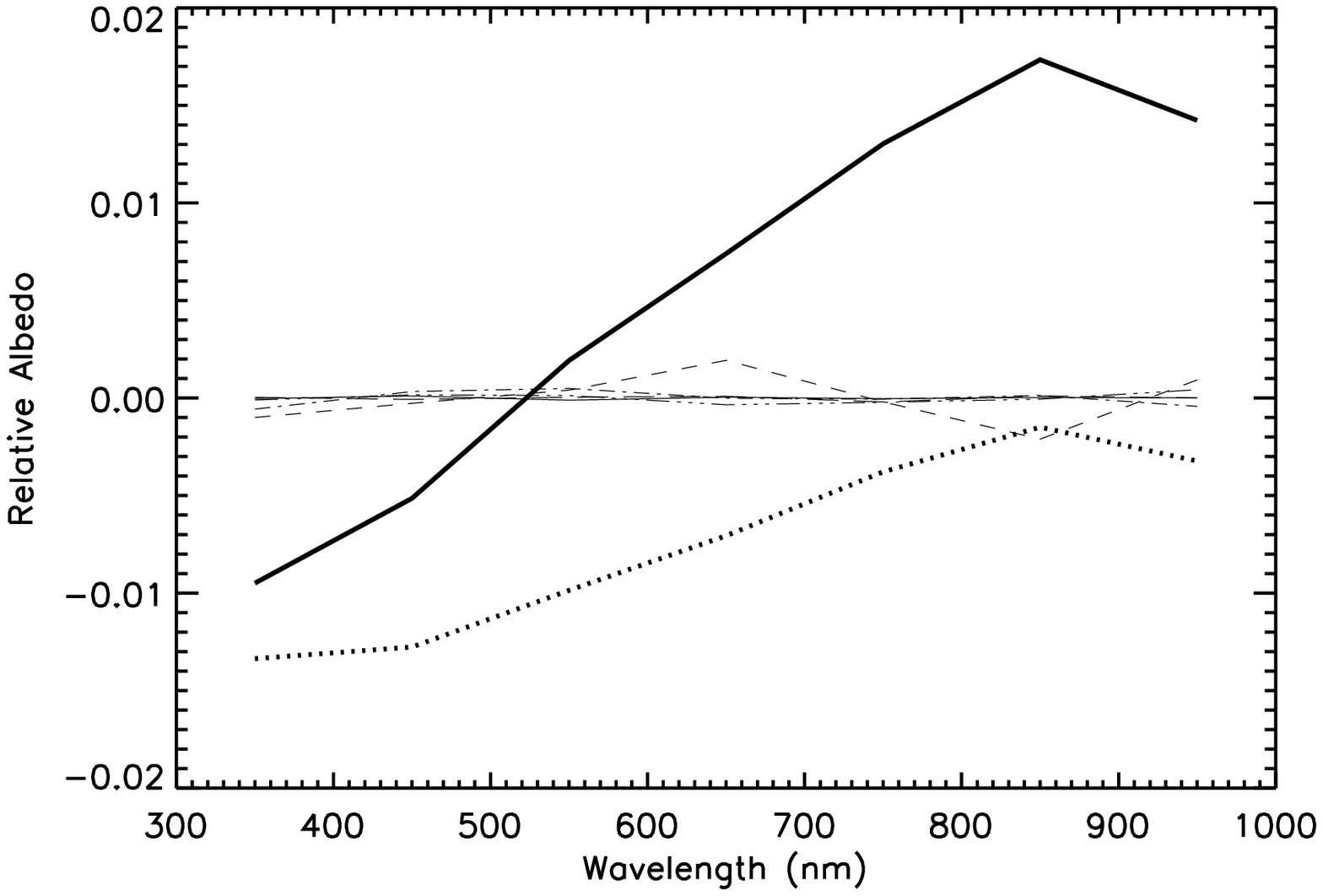}
\includegraphics[width=84mm]{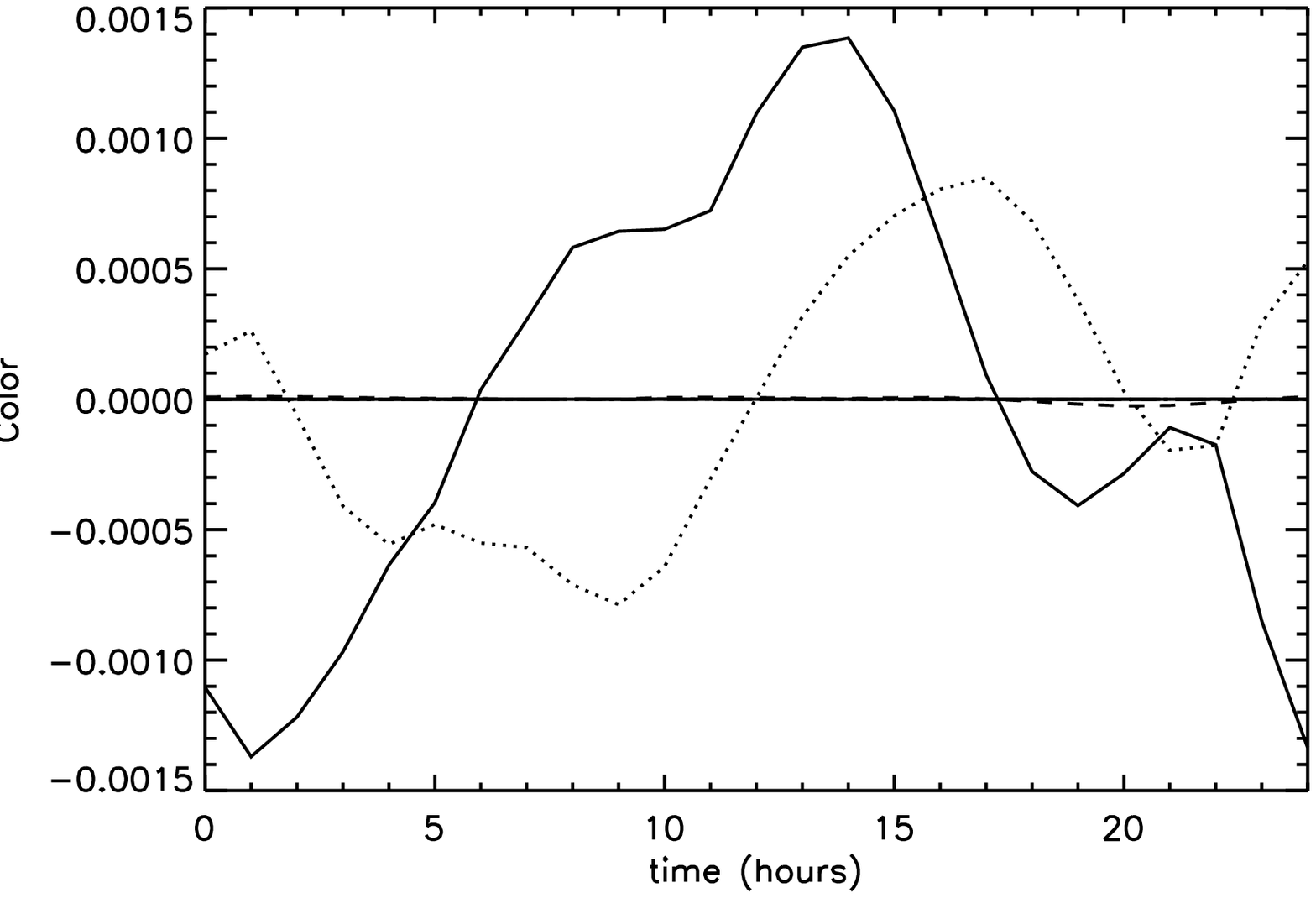}
\caption{{\bf Earth1 EPOXI Observations} \emph{Top Left:} Time-averaged broadband spectrum. \emph{Top Right:} Normalized variability spectrum from PCA. \emph{Bottom Left:} Eigencolors from PCA. The eigenspectra have been normalized by their eigenvalues, so the dominant components exhibit larger excursions from zero. \emph{Bottom Right:} Eigenprojections from PCA.}
\label{Earth1_all}
\end{figure*}

In Figure~\ref{Earth1_all} we summarize the results of the PCA performed on the 2008 EPOXI Equinox data. These are the same as the results presented in \cite{Cowan_2009}: the dominant eigencolor is red (most non-zero at long wavelengths), while the second eigencolor is blue (most non-zero at short wavelengths). Note that the sign of the eigenspectra, and hence its slope, is not important in describing its color. Although the two primary eigencolors shown in Figure~\ref{Earth1_all} look similar at first glance, they are in fact orthogonal, by definition. 
     
We now run five different versions of the VPL 3D Earth model: 1) Standard: this model is an excellent fit to the EPOXI Earth1 observations; the remaining models are identical, but in each case a single model element has been ``turned off'': 2) Cloud Free; 3) No Rayleigh Scattering; 4) Black Oceans; 5) Black Land.

\begin{figure*}[htb]
\includegraphics[width=84mm]{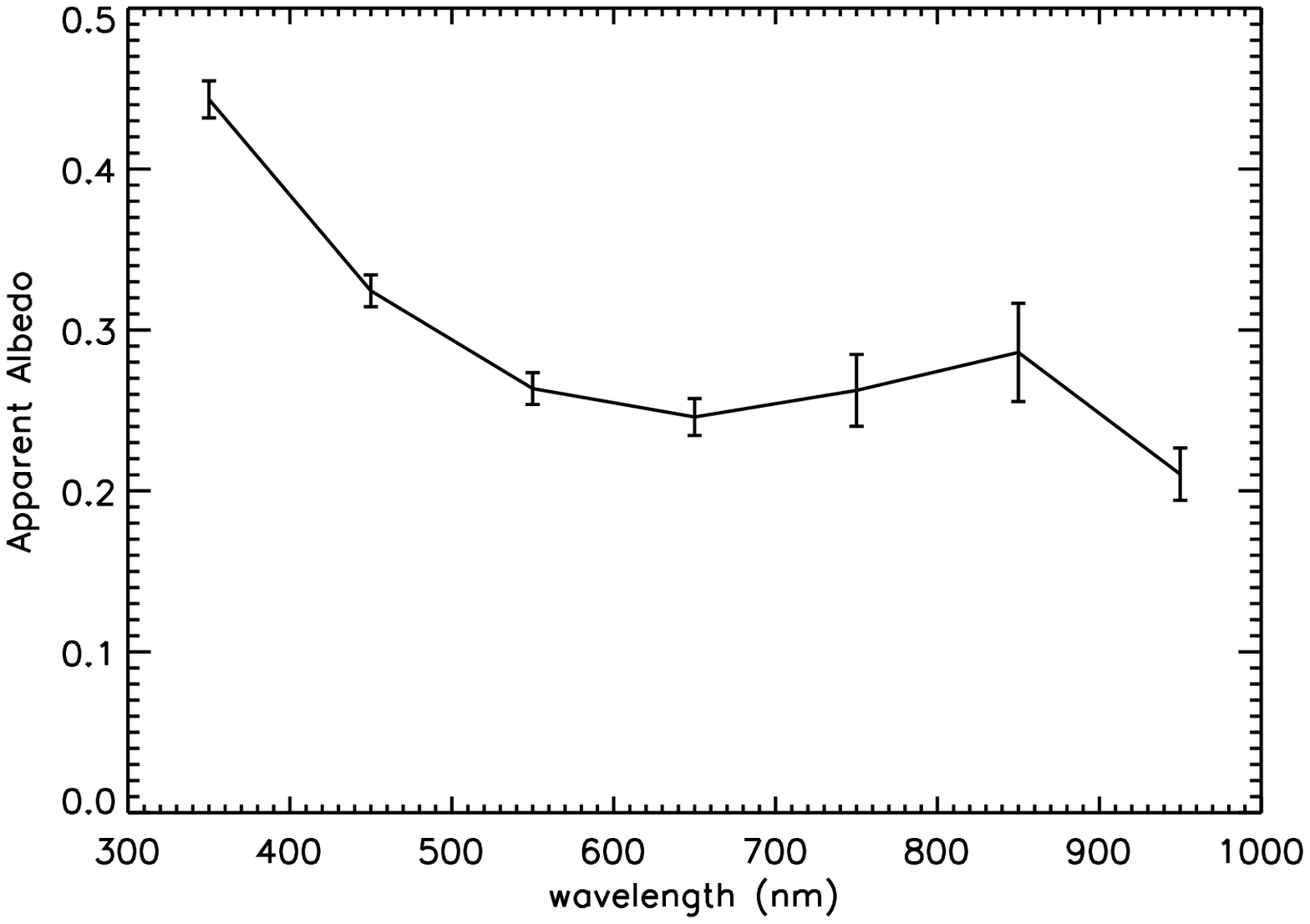}
\includegraphics[width=84mm]{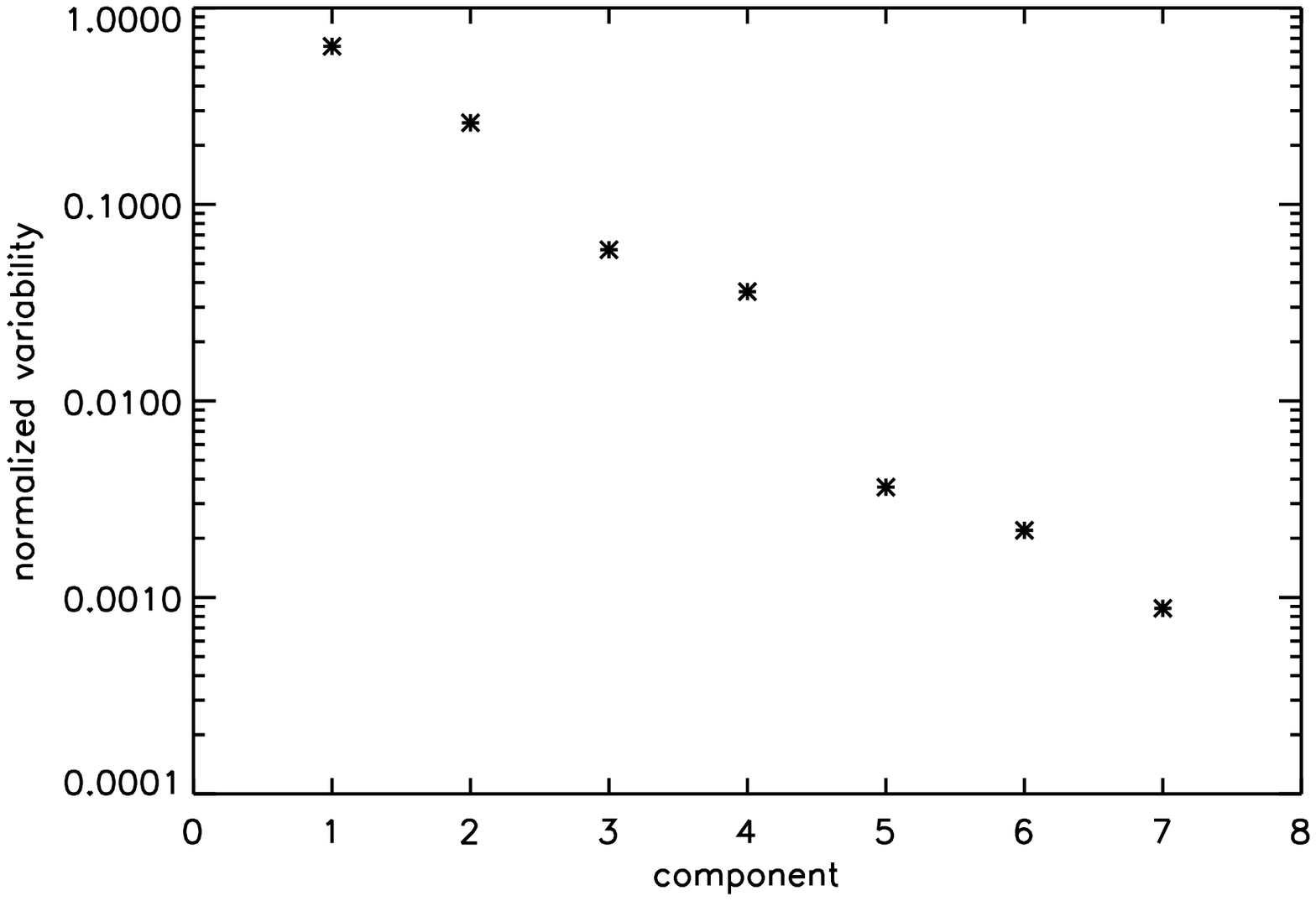}
\includegraphics[width=84mm]{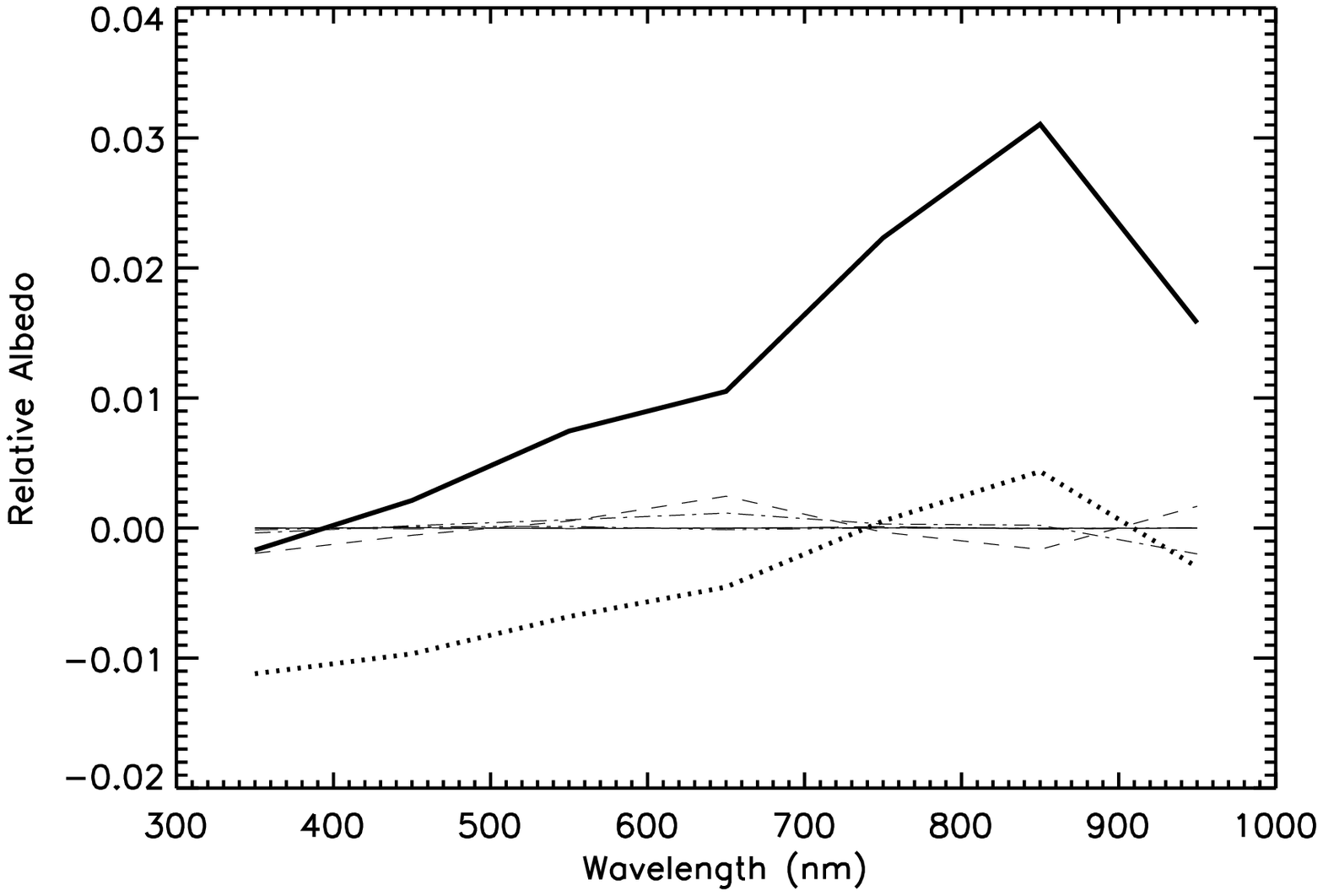}
\includegraphics[width=84mm]{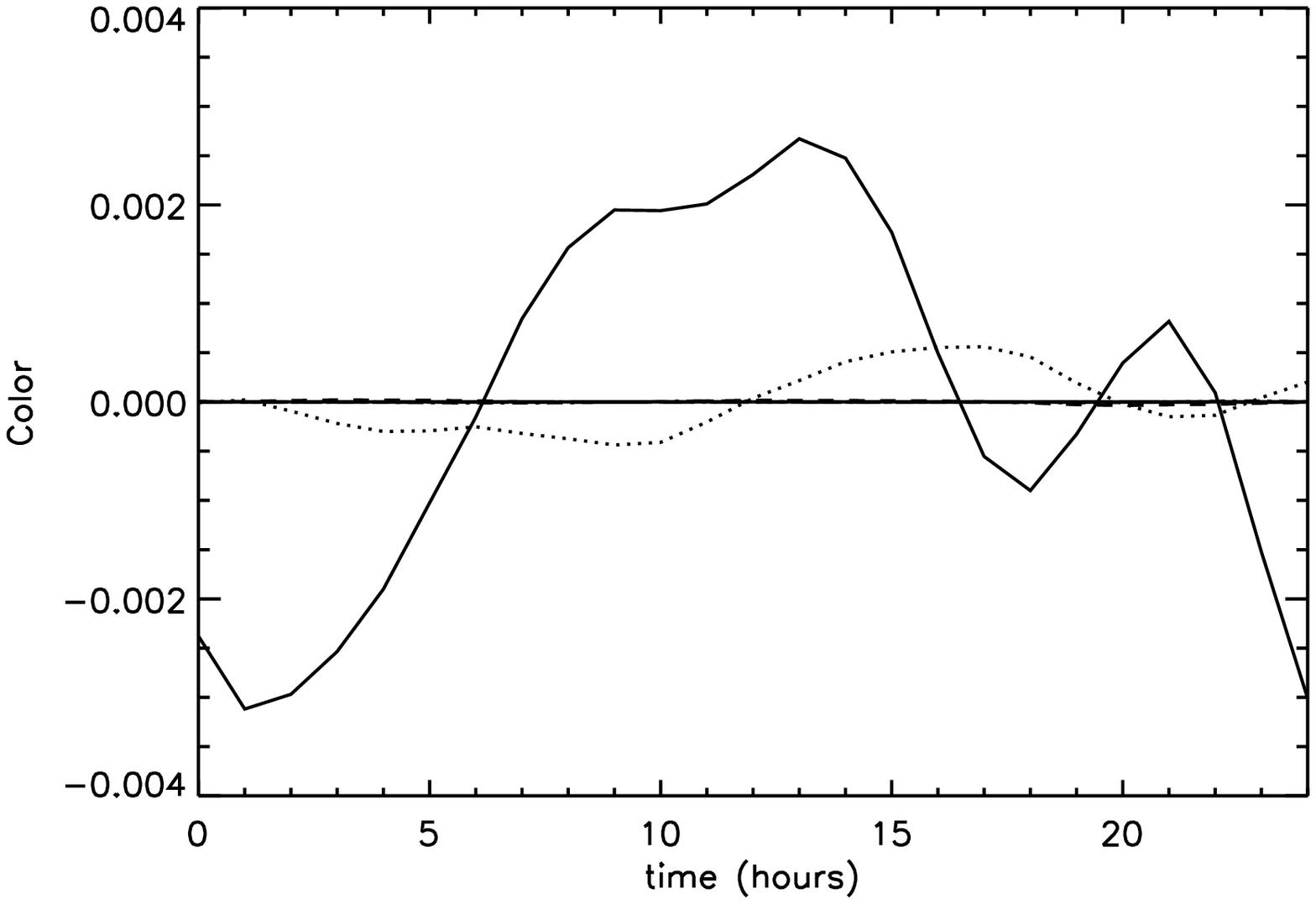}
\caption{{\bf Standard VPL Simulation} \emph{Top Left:} Time-averaged broadband spectrum. \emph{Top Right:} Normalized variability spectrum from PCA. \emph{Bottom Left:} Eigencolors from PCA. The eigenspectra have been normalized by their eigenvalues, so the dominant components exhibit larger excursions from zero. \emph{Bottom Right:} Eigenprojections from PCA.}
\label{standard_all}
\end{figure*}

1) The Standard model (Figure~\ref{standard_all}) produces eigencolors indistinguishable from those presented in \cite{Cowan_2009} or the control case above: a dominant red eigencolor followed closely by a blue eigencolor. The relative importance of the eigencolors as a function of time, ``eigenprojections'', also match very well. This should not be surprising, given the excellent fit to the actual data (Robinson et al. 2011).  

\begin{figure*}[htb]
\includegraphics[width=84mm]{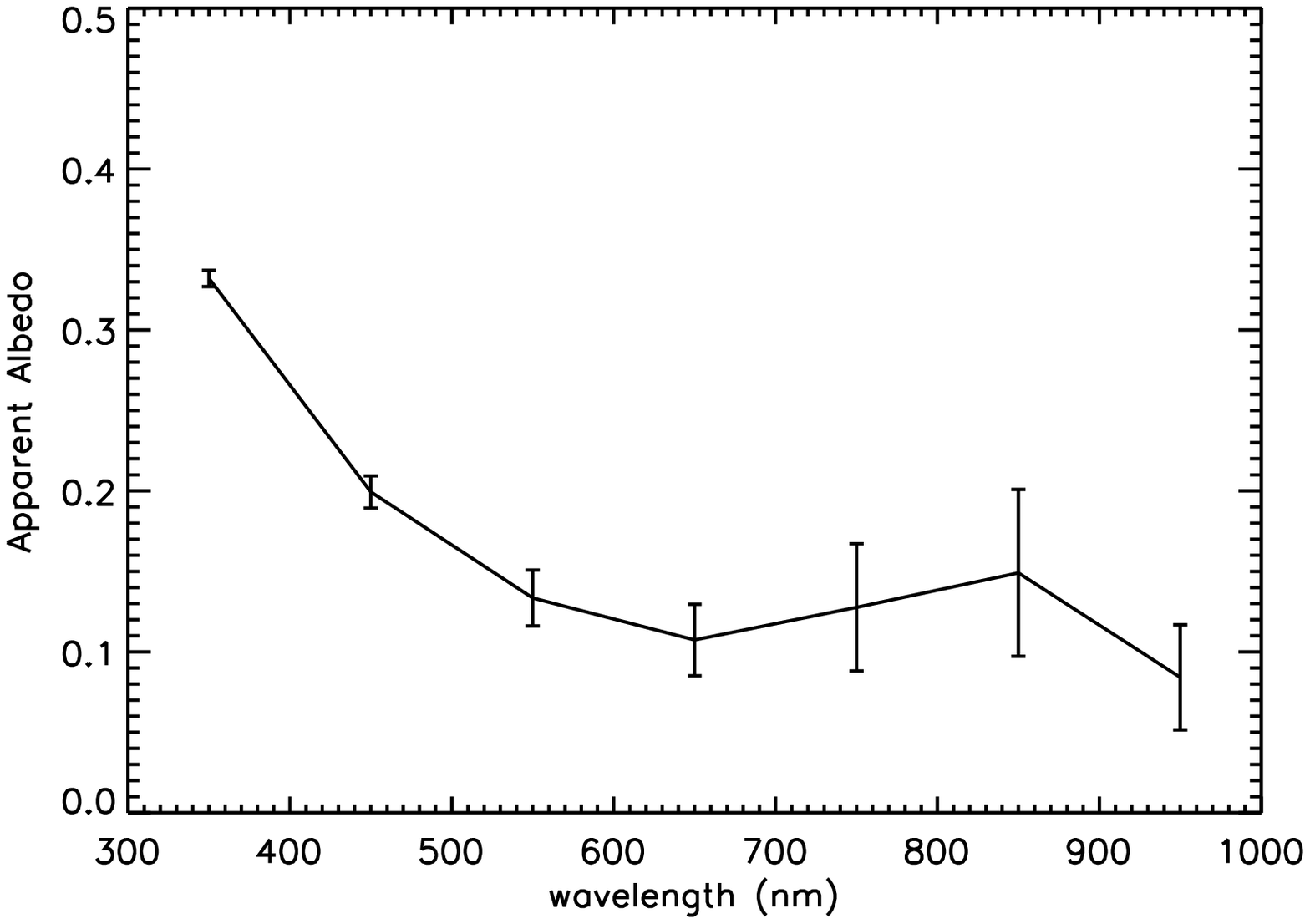}
\includegraphics[width=84mm]{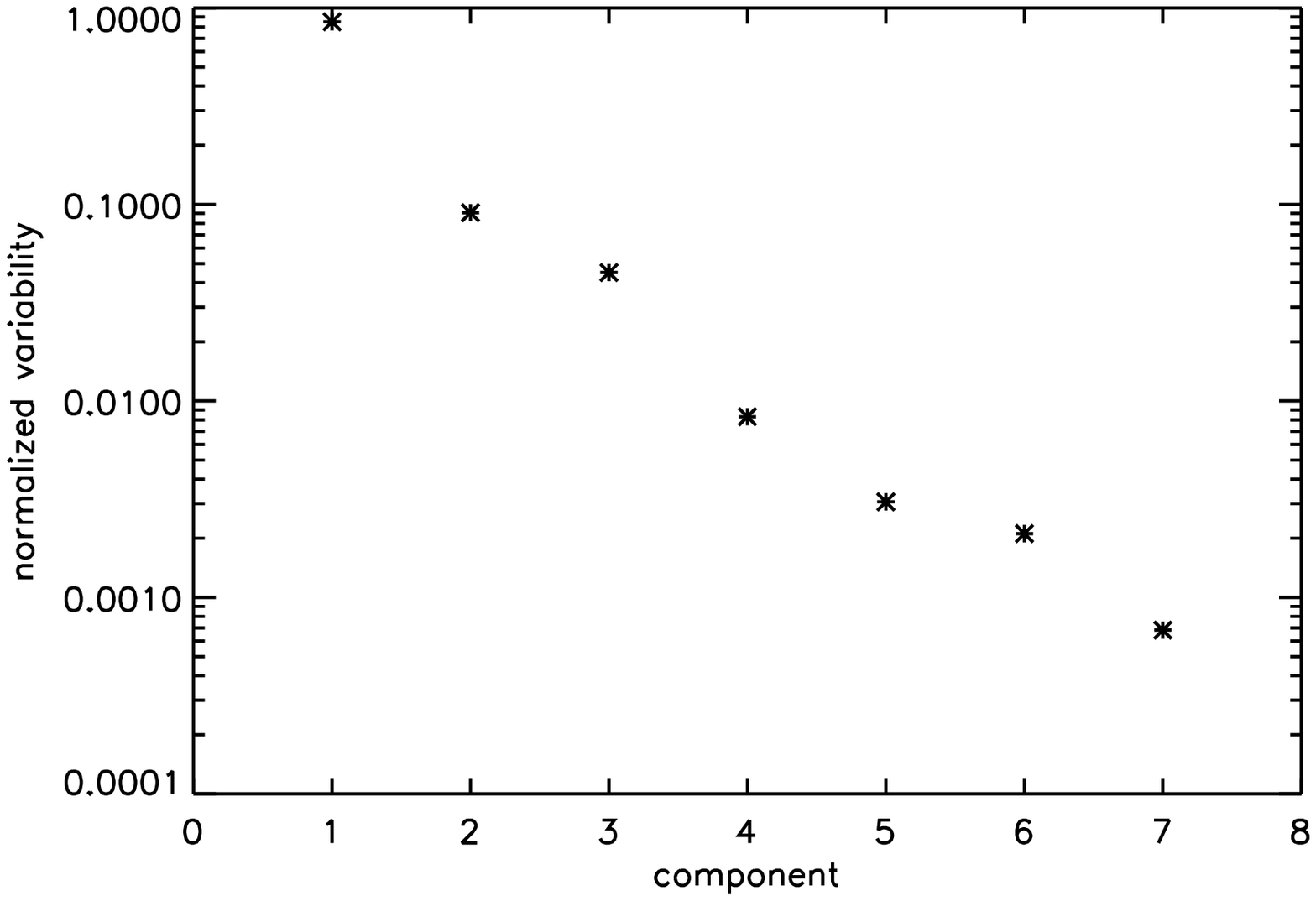}
\includegraphics[width=84mm]{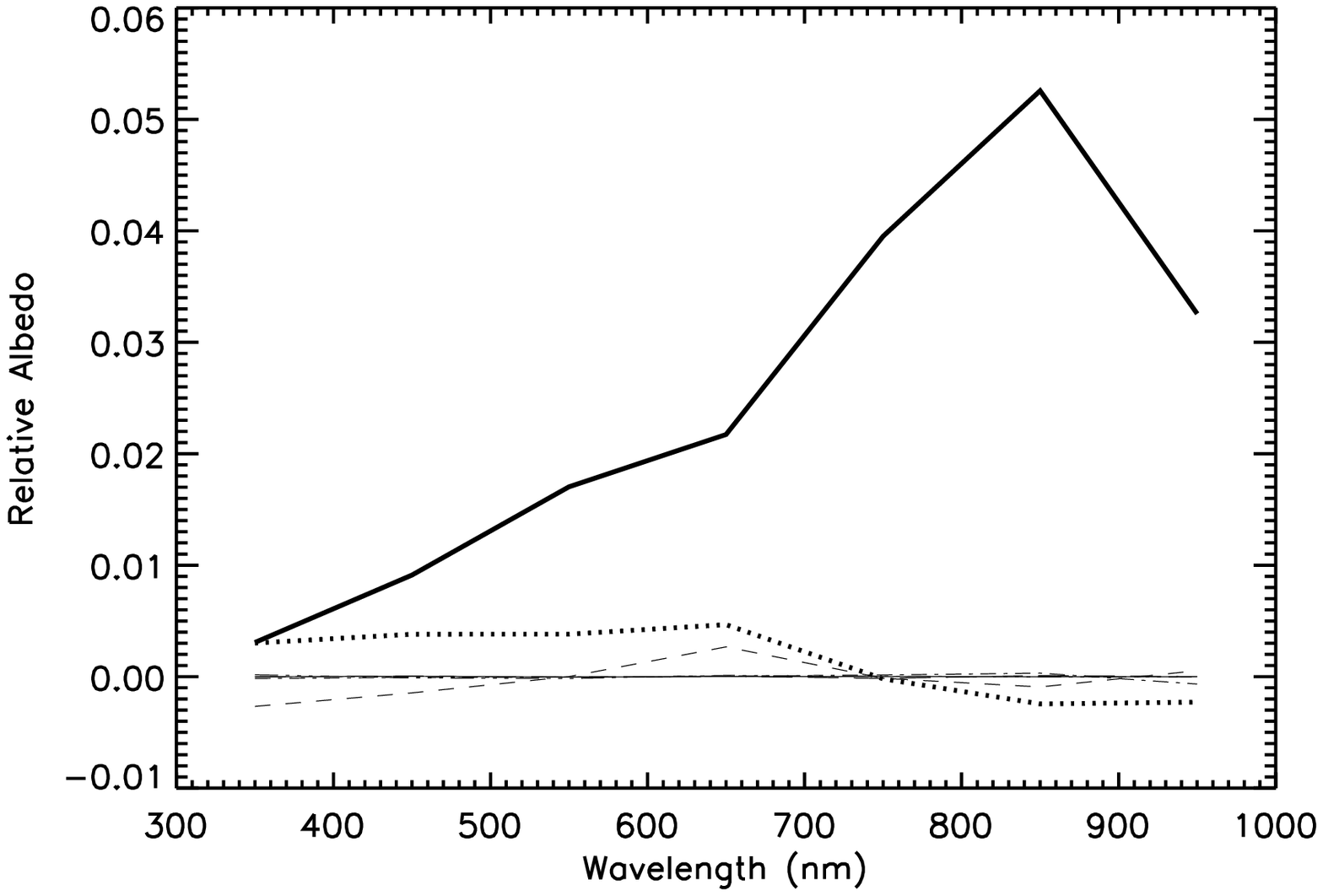}
\includegraphics[width=84mm]{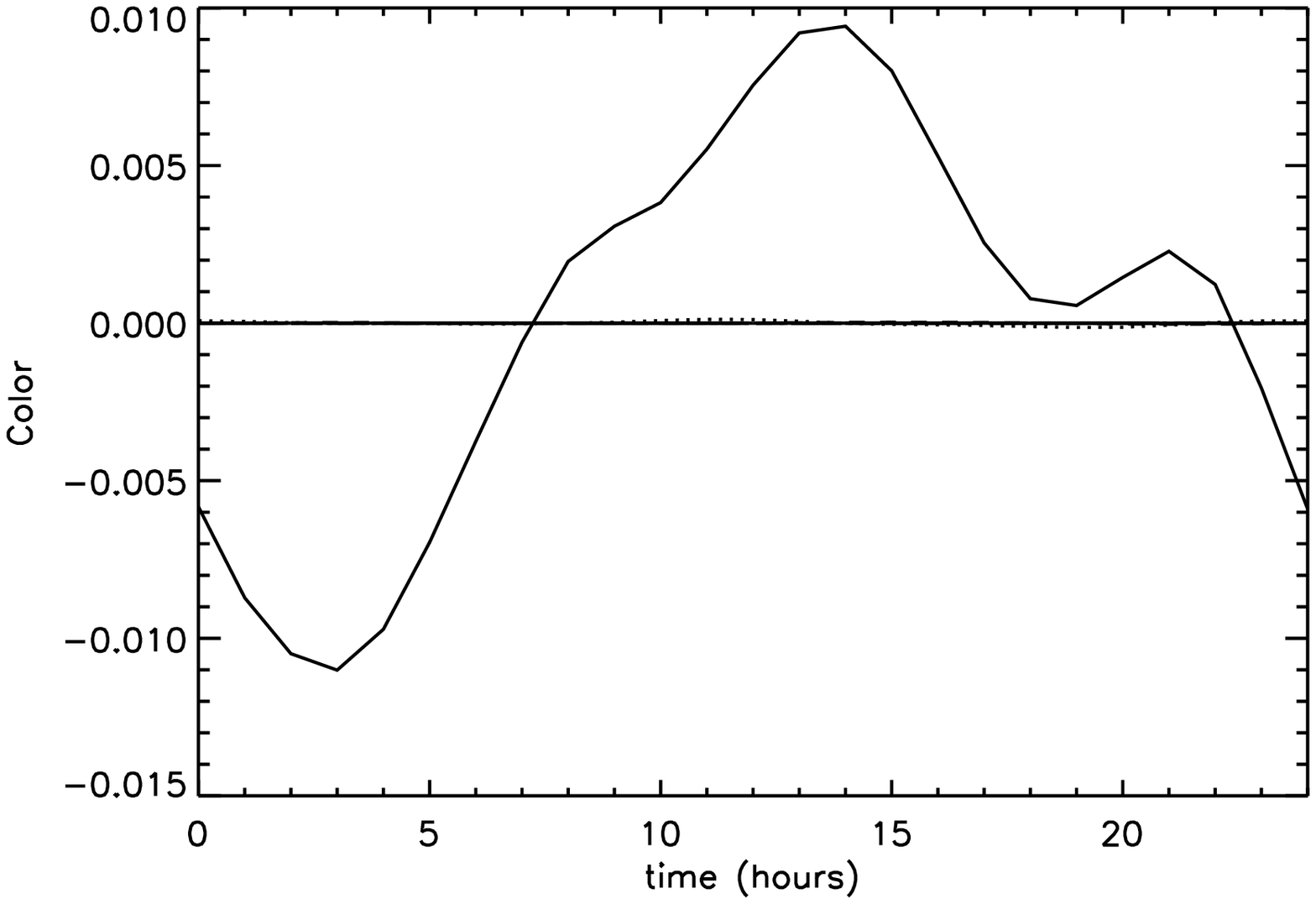}
\caption{{\bf Cloud Free VPL Simulation} \emph{Top Left:} Time-averaged broadband spectrum. \emph{Top Right:} Normalized variability spectrum from PCA. \emph{Bottom Left:} Eigencolors from PCA. The eigenspectra have been normalized by their eigenvalues, so the dominant components exhibit larger excursions from zero. \emph{Bottom Right:} Eigenprojections from PCA.}
\label{no_cloud_all}
\end{figure*}

2) The Cloud Free model (Figure~\ref{no_cloud_all}) has similar time-averaged colors to the Standard model, but is less reflective at all wavelengths ($\Delta A^{*}\approx -0.1$).  This is especially noticeable at long wavelengths, where Rayleigh scattering does not operate. If the albedo were not on an absolute scale, as would be the case for a directly-imaged planet with no reliable radius estimate, it would be difficult to distinguish this cloud-free planet from its cloudy counterpart. Unlike the Standard case, however, the Cloud Free model shows very little variability at blue wavebands.  As a result, the Cloud Free model shows the same dominant red eigencolor as the Standard model, but the amplitude of excursions for the blue eigencolor are much smaller than for the Standard model.

\begin{figure*}[htb]
\includegraphics[width=84mm]{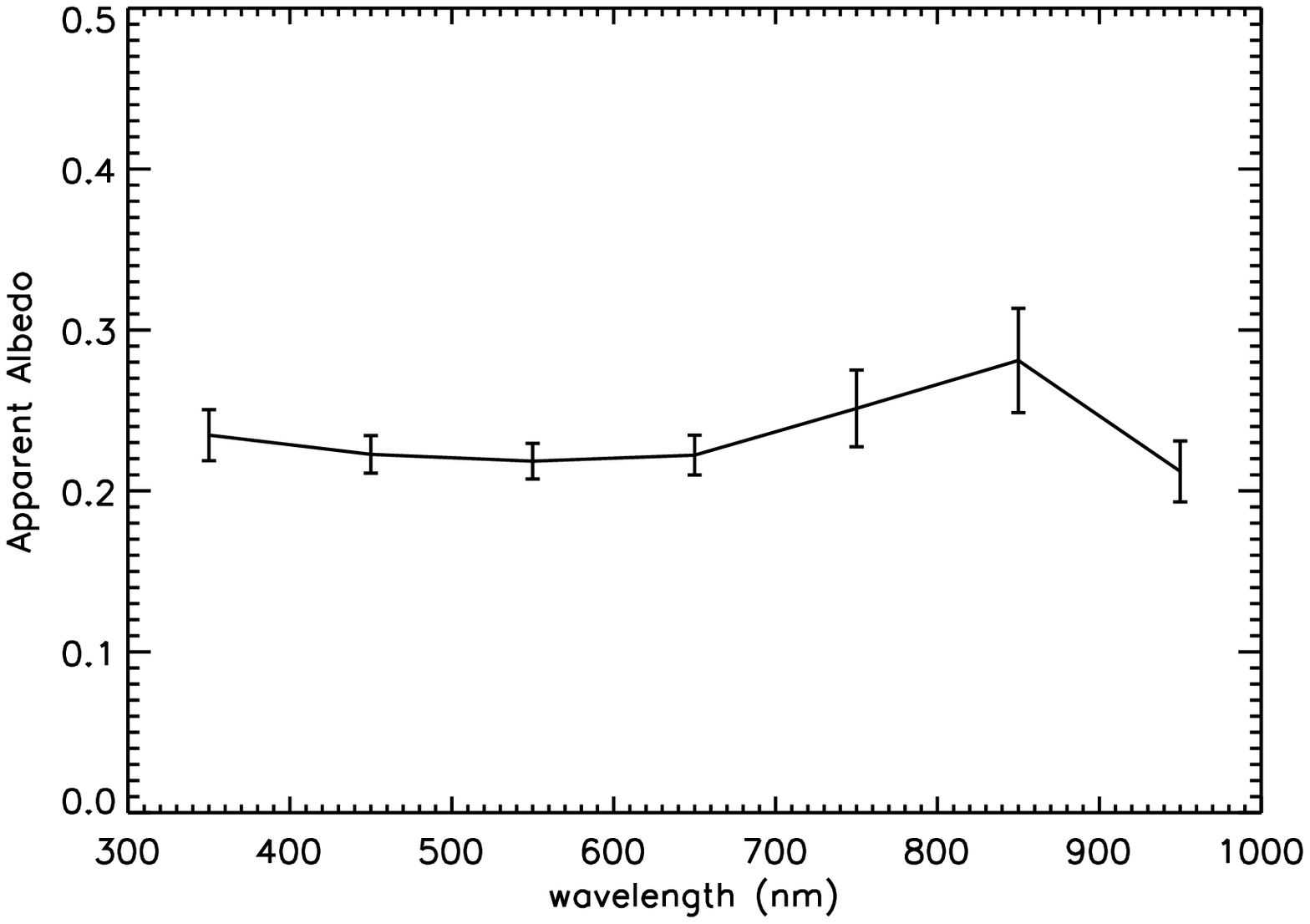}
\includegraphics[width=84mm]{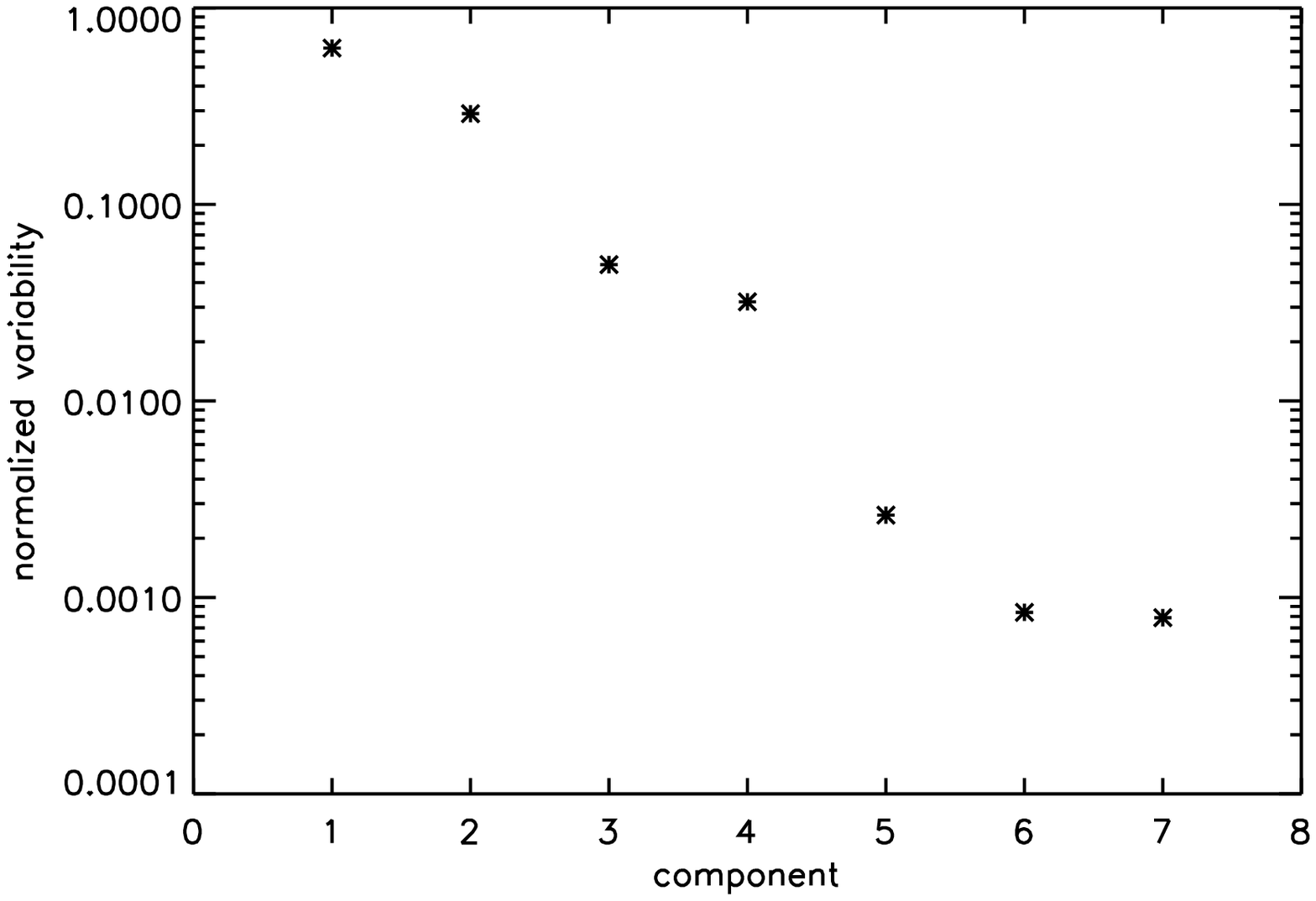}
\includegraphics[width=84mm]{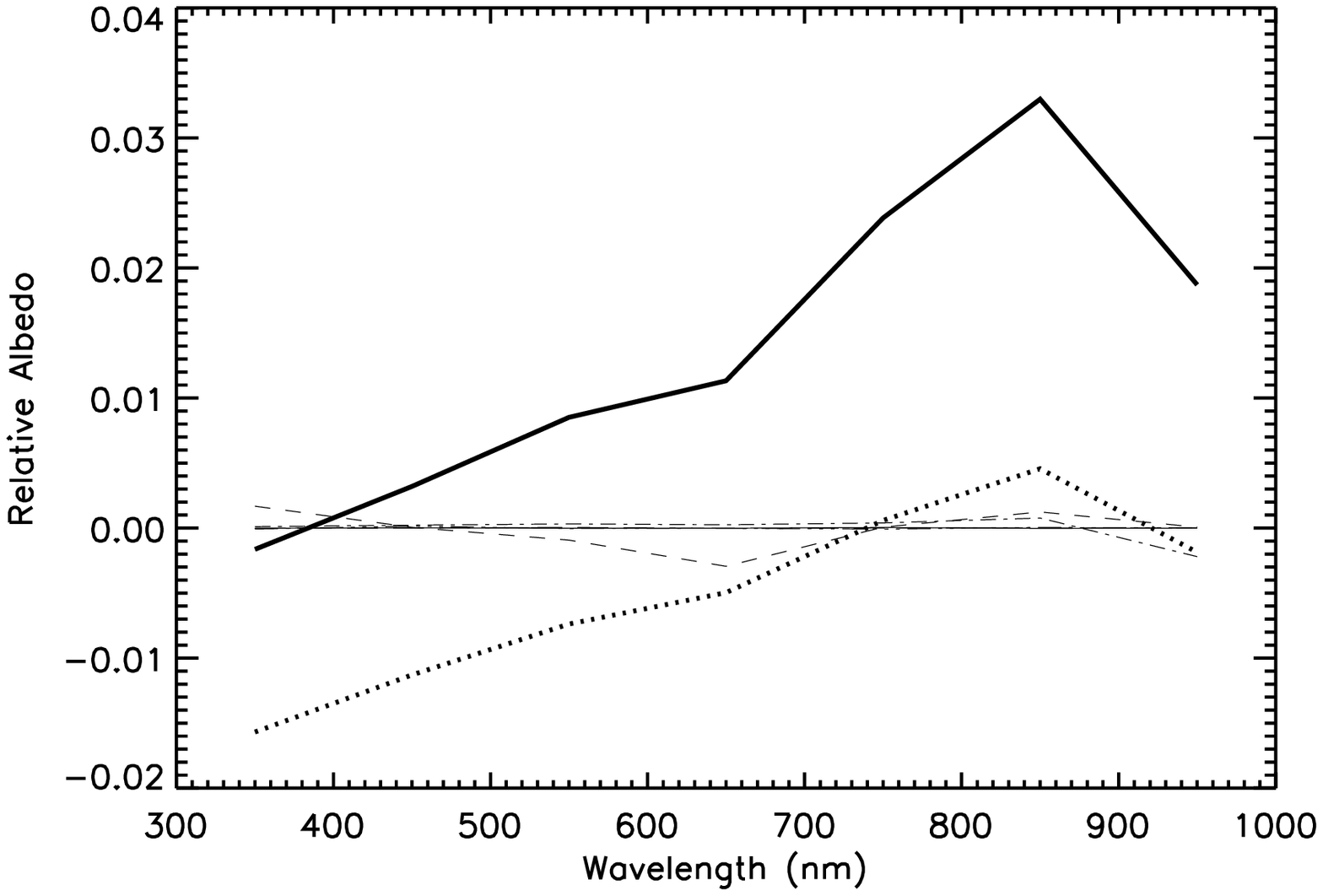}
\includegraphics[width=84mm]{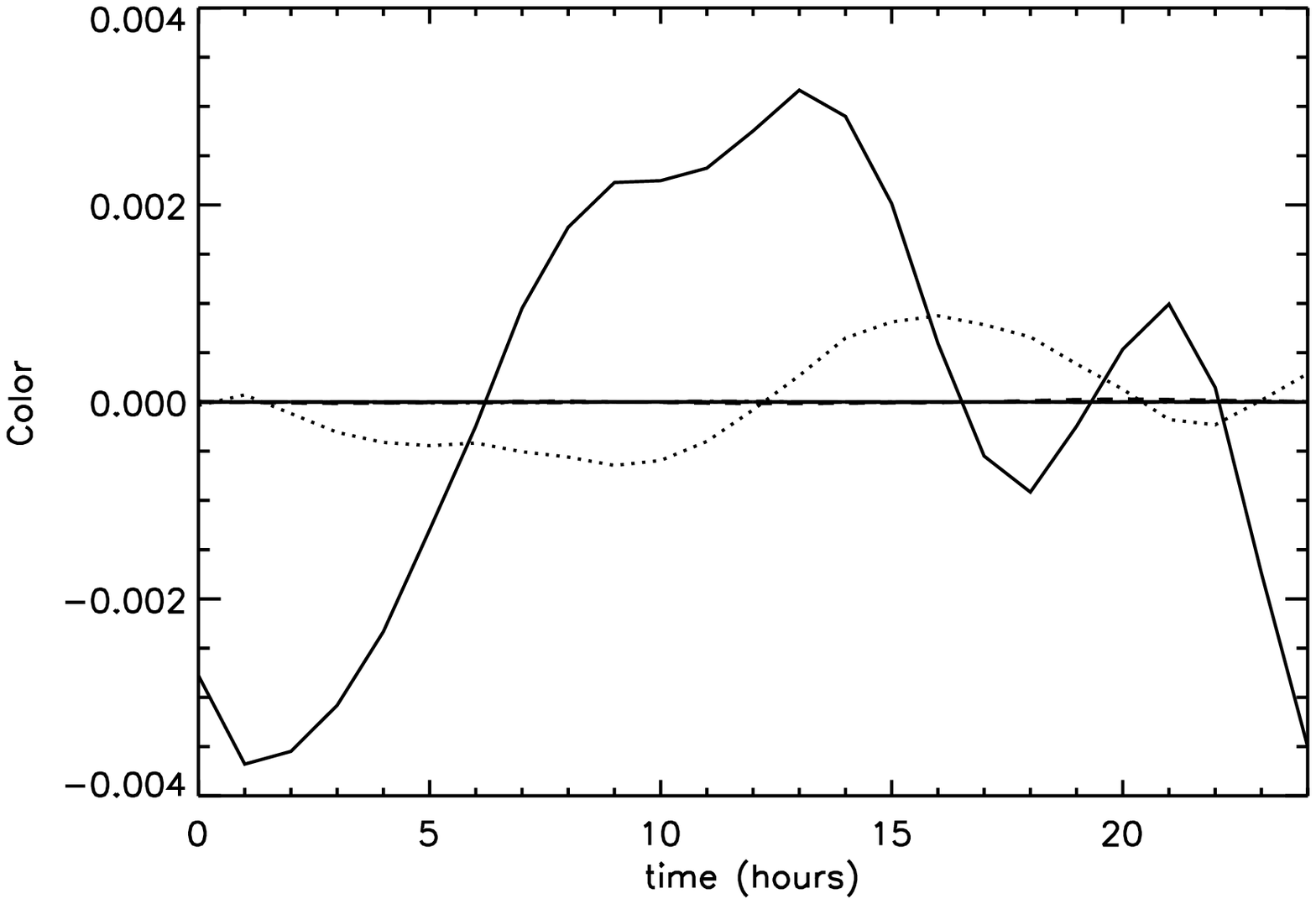}
\caption{{\bf No Rayleigh Scattering VPL Simulation} \emph{Top Left:} Time-averaged broadband spectrum. \emph{Top Right:} Normalized variability spectrum from PCA. \emph{Bottom Left:} Eigencolors from PCA. The eigenspectra have been normalized by their eigenvalues, so the dominant components exhibit larger excursions from zero. \emph{Bottom Right:} Eigenprojections from PCA.}
\label{no_rayleigh_all}
\end{figure*}

3) The No Rayleigh Scattering model (Figure~\ref{no_rayleigh_all}) has red time-averaged colors, with a slight upturn in reflectance at the bluest wavebands due to oceans.  The eigencolors and eigenprojections are essentially the same as in the Standard model.  

\begin{figure*}[htb]
\includegraphics[width=84mm]{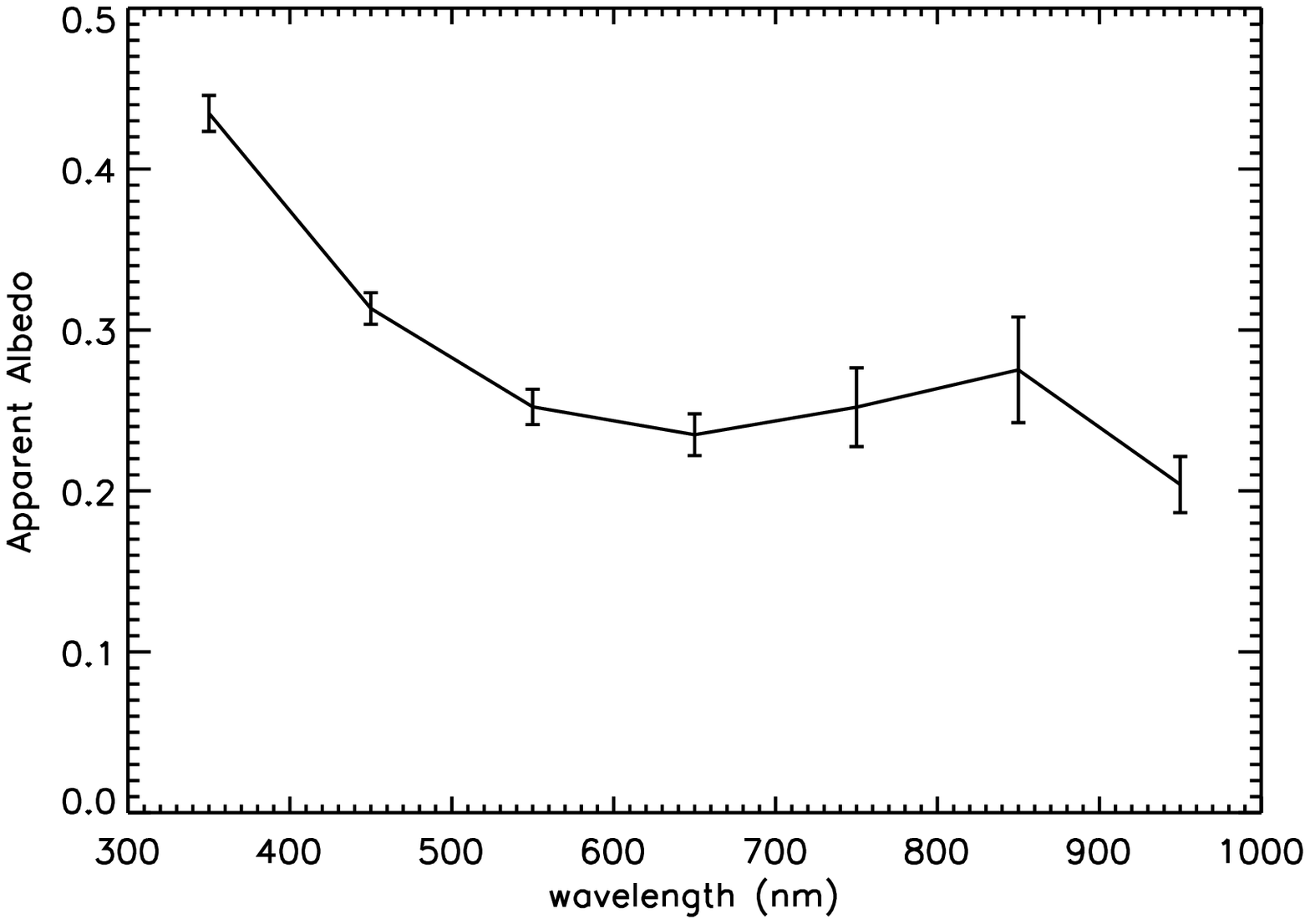}
\includegraphics[width=84mm]{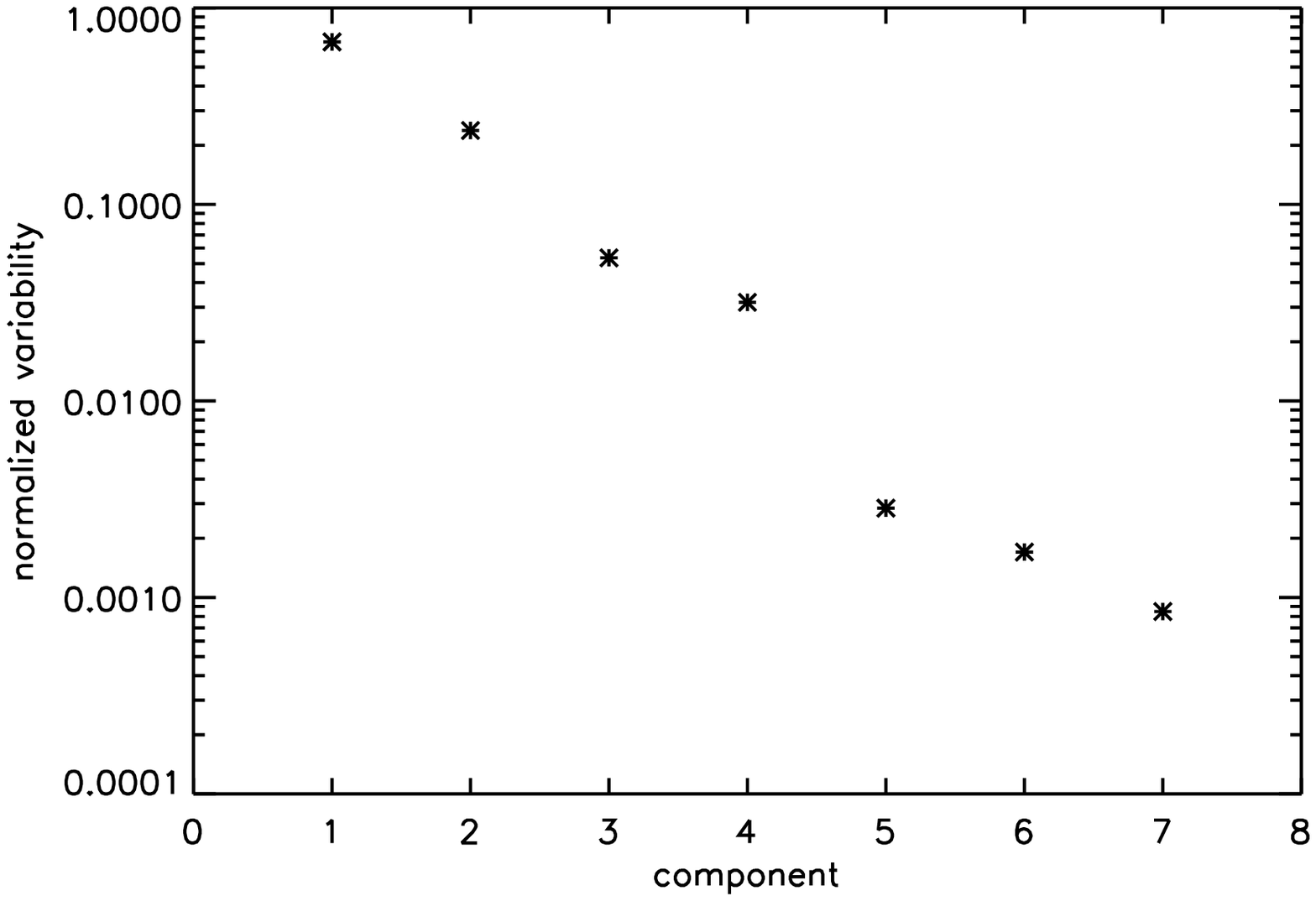}
\includegraphics[width=84mm]{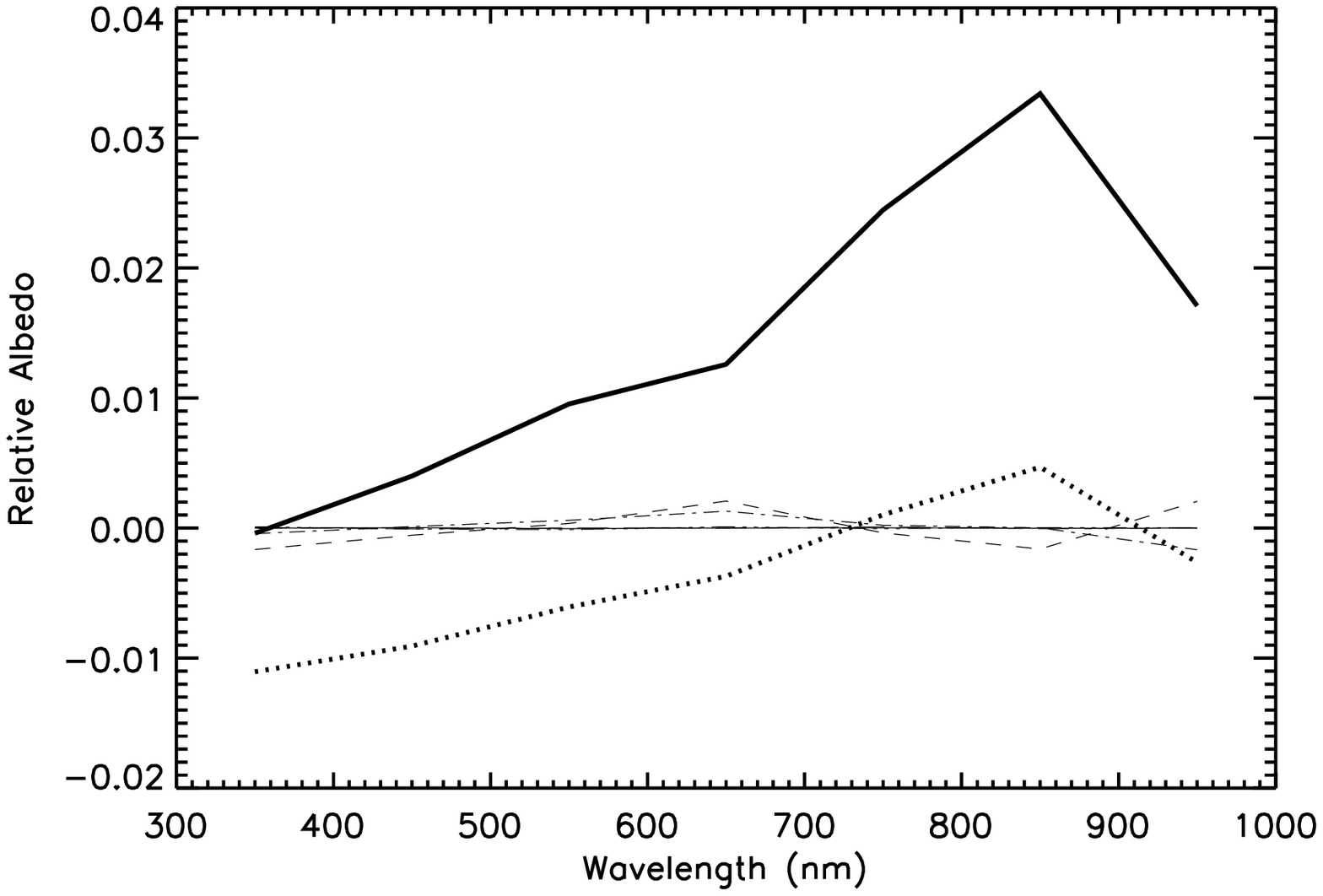}
\includegraphics[width=84mm]{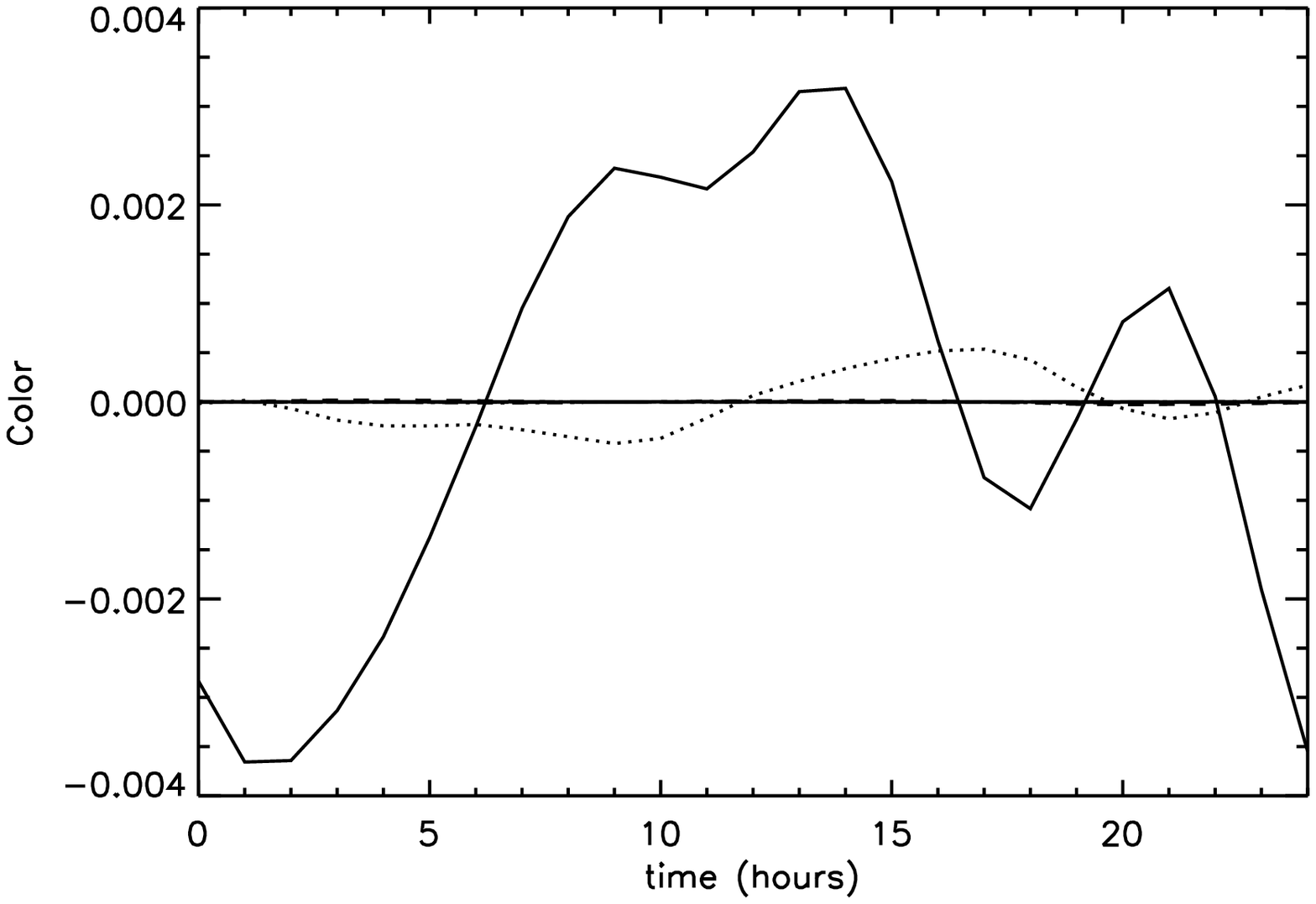}
\caption{{\bf Black Oceans VPL Simulation} \emph{Top Left:} Time-averaged broadband spectrum. \emph{Top Right:} Normalized variability spectrum from PCA. \emph{Bottom Left:} Eigencolors from PCA. The eigenspectra have been normalized by their eigenvalues, so the dominant components exhibit larger excursions from zero. \emph{Bottom Right:} Eigenprojections from PCA.}
\label{zero_albedo_ocean_all}
\end{figure*}

4) The Black Oceans model (Figure~\ref{zero_albedo_ocean_all}) has time-averaged colors, eigencolors and eigenprojections indistinguishable from those of the Standard model. This indicates that at gibbous phases oceans on Earth consist of a null surface type, contributing neither to the time-averaged nor to the time-resolved disk-integrated colors. This does not preclude, however, the importance of specular reflection at crescent phases. 

\begin{figure*}[htb]
\includegraphics[width=84mm]{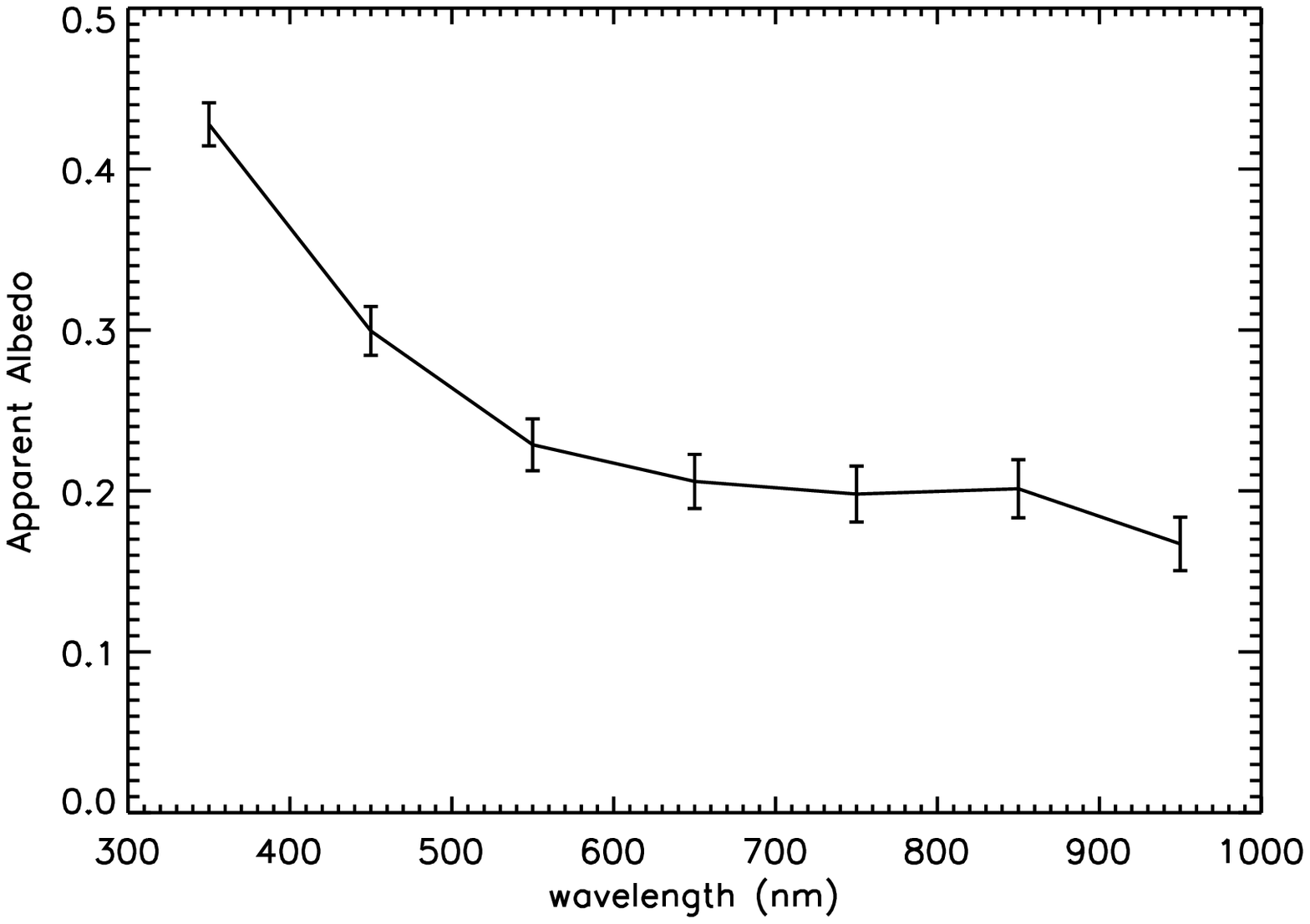}
\includegraphics[width=84mm]{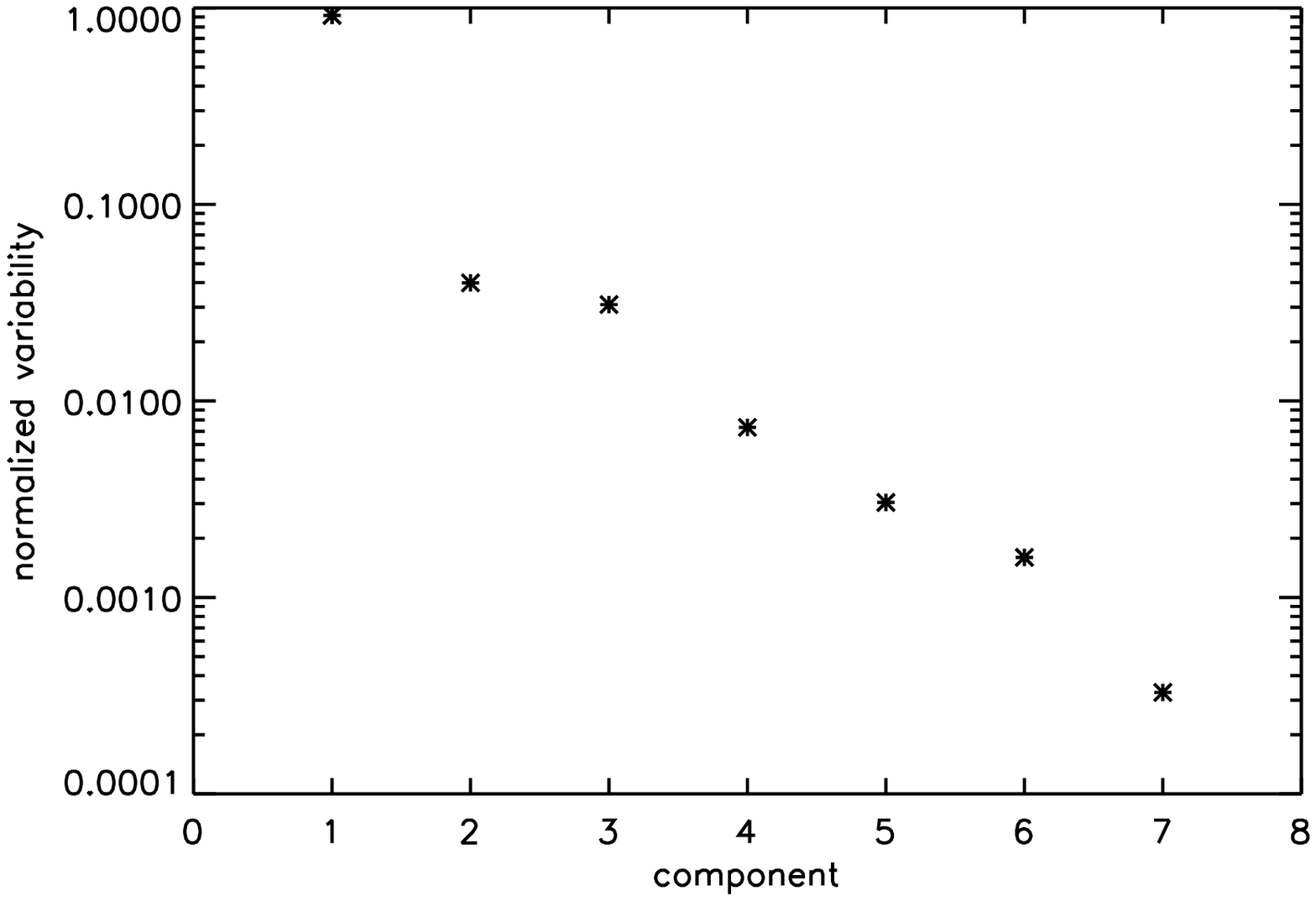}
\includegraphics[width=84mm]{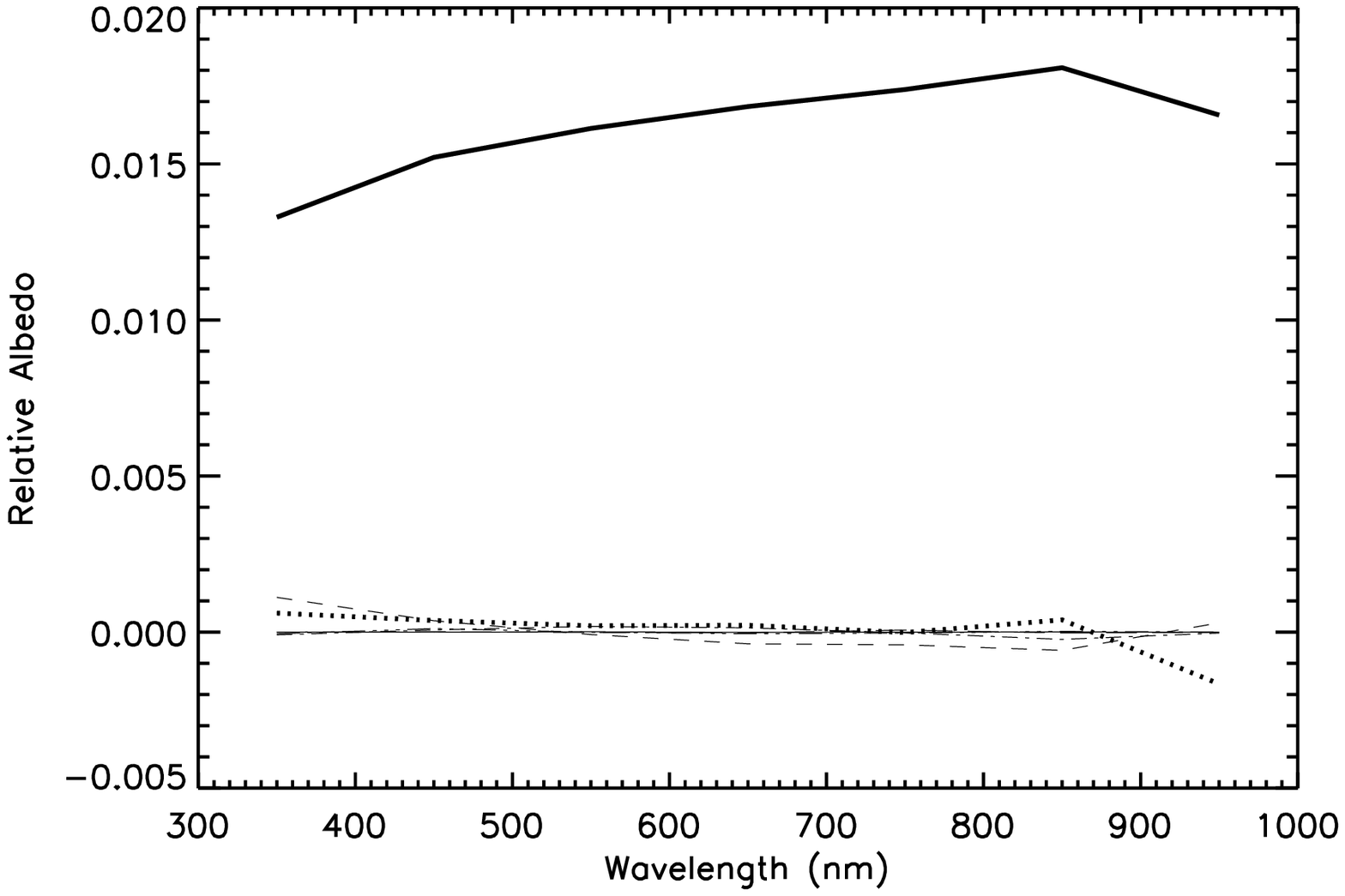}
\includegraphics[width=84mm]{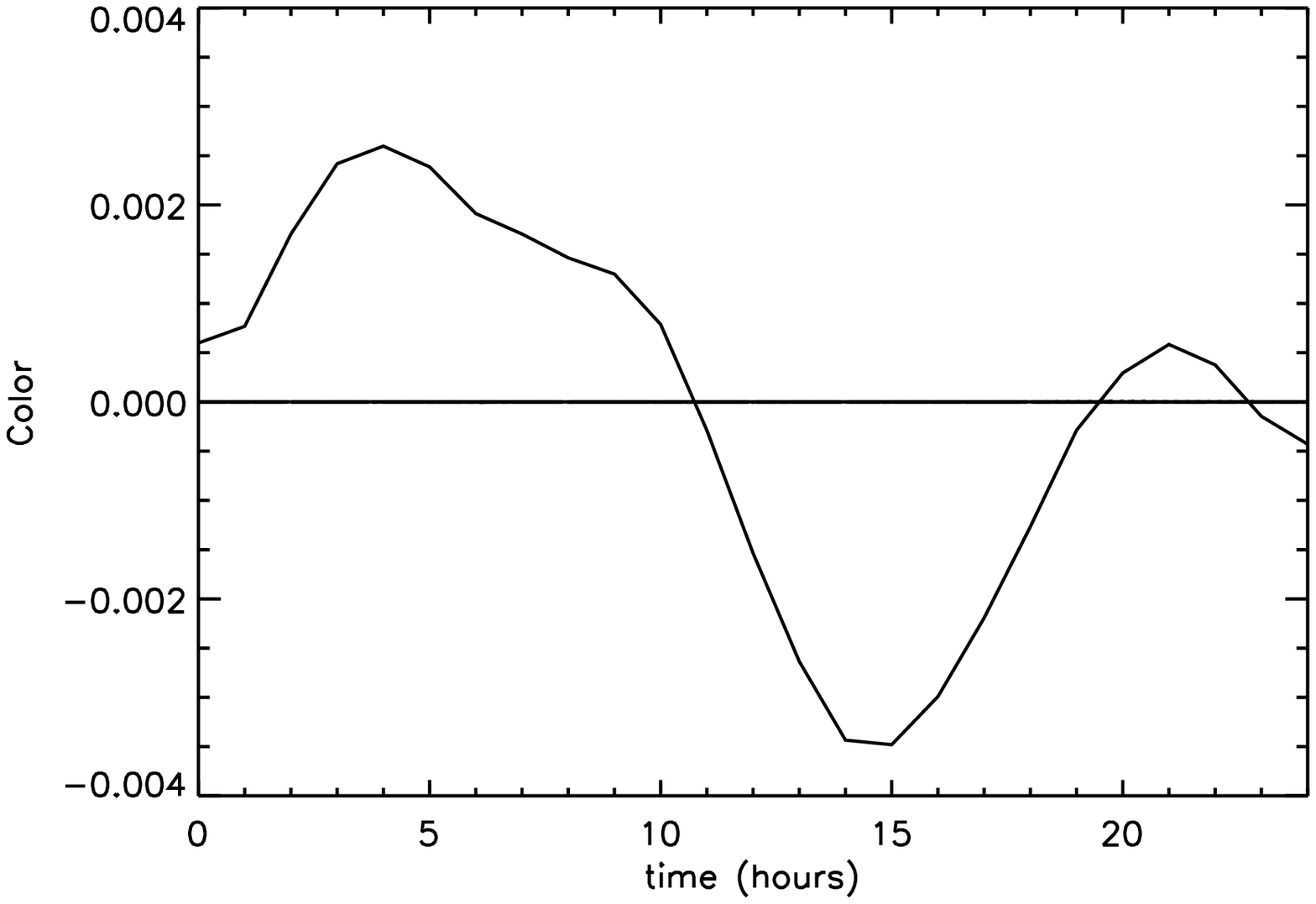}
\caption{{\bf Black Land VPL Simulation} \emph{Top Left:} Time-averaged broadband spectrum. \emph{Top Right:} Normalized variability spectrum from PCA. \emph{Bottom Left:} Eigencolors from PCA. The eigenspectra have been normalized by their eigenvalues, so the dominant components exhibit larger excursions from zero. \emph{Bottom Right:} Eigenprojections from PCA.}
\label{zero_albedo_land_all}
\end{figure*}

5) The Black Land model (Figure~\ref{zero_albedo_land_all}) has similar time-averaged colors to the Standard model, but without the upturn at near-IR wavelengths. There is a single dominant, gray eigencolor.

\end{document}